\documentclass[lettersize,journal]{IEEEtran}
\usepackage{amsmath,amsfonts}
\usepackage{array}
\usepackage{textcomp}
\usepackage{stfloats}
\usepackage{url}
\usepackage{verbatim}
\usepackage{graphicx}
\usepackage{cite}
\usepackage{xcolor}
\usepackage {xspace}
\usepackage{booktabs}
\usepackage{algpseudocode} 
\usepackage{mathrsfs}
\usepackage{subcaption}
\usepackage{tabu}                 
\usepackage{multirow}             
\usepackage{multicol}             
\usepackage{multirow}             
\usepackage{float}                
\usepackage{makecell}             
\usepackage{booktabs}             

\usepackage[ruled,norelsize,linesnumbered]{algorithm2e}
\newcommand*{\note}[1]{\textcolor{red}{#1}}
\newcommand*{\cpy}[1]{\textcolor{black}{#1}}
\newcommand*{\newinfo}[1]{\textcolor{black}{#1}}
\newcommand*{\newnew}[1]{\textcolor{black}{#1}}
\newcommand{\algname}{HG-DRL\xspace}

\DeclareMathOperator*{\concat}{%
    \mathchoice%
        {\Big\Vert}%
        {\big\Vert}%
        {\Vert}%
        {\Vert}%
}

\begin{document}

\title{Towards Multi-agent Reinforcement Learning based Traffic Signal Control through Spatio-temporal Hypergraphs}







\author{Kang Wang,
        Zhishu Shen,~\IEEEmembership{Member,~IEEE},
        Zhen Lei, 
        Xianhui Liu and
        Tiehua Zhang,~\IEEEmembership{Member,~IEEE}

\thanks{ \textit{Corresponding author: Zhishu Shen and Tiehua Zhang}}
\thanks{Kang Wang, Zhishu Shen and Zhen Lei are with the School of Computer Science and Artificial Intelligence, Wuhan University of Technology, Wuhan, China (e-mail: wangk7733@whut.edu.cn, z\_shen@ieee.org, studentlz@whut.edu.cn). Zhishu Shen is also with the Hubei Key Laboratory of Transportation Internet of Things,  School of Computer Science and Artificial Intelligence, Wuhan University of Technology, Wuhan, China.}

\thanks{Xianhui Liu is with the College of Electronics and Information Engineering, Tongji University, Shanghai, China (e-mail: lxh@tongji.edu.cn).}

\thanks{Tiehua Zhang is with the School of Computer Science and Technology, Tongji University, Shanghai, China (e-mail: tiehuaz@tongji.edu.cn).}

\thanks{This work was supported in part by the National Natural Science Foundation of China (Grant No. 62472332 and 62372330).} 

}

\markboth{Journal of \LaTeX\ Class Files,~Vol.~14, No.~8, August~2021}%
{Shell \MakeLowercase{\textit{et al.}}: A Sample Article Using IEEEtran.cls for IEEE Journals}



\maketitle

\begin{abstract}
 Traffic signal control systems (TSCSs) are integral to intelligent traffic management, fostering efficient vehicle flow. Traditional approaches often simplify road networks into standard graphs, which results in a failure to consider the dynamic nature of traffic data at neighboring intersections, thereby neglecting higher-order interconnections necessary for real-time control. To address this, we propose a novel TSCS framework to realize intelligent traffic control. This framework collaborates with multiple neighboring edge computing servers to collect traffic information across the road network. To elevate the efficiency of traffic signal control, we have crafted a multi-agent soft actor-critic (MA-SAC) reinforcement learning algorithm. Within this algorithm, individual agents are deployed at each intersection with a mandate to optimize traffic flow across the road network collectively. Furthermore, we introduce hypergraph learning into the critic network of MA-SAC to enable the spatio-temporal interactions from multiple intersections in the road network. This method fuses hypergraph and spatio-temporal graph structures to encode traffic data and capture the complex spatio-temporal correlations between multiple intersections. Our empirical evaluation, tested on varied datasets, demonstrates the superiority of our framework in minimizing average vehicle travel times and sustaining high-throughput performance. This work facilitates the development of more intelligent urban traffic management solutions. \newnew{We release the code to support the reproducibility of this work at https://github.com/Edun-Eyes/TSC}.
 
\end{abstract}

\begin{IEEEkeywords}
Traffic signal control, mobile data mining, multi-access edge computing, hypergraph learning, deep reinforcement learning
\end{IEEEkeywords}

\section{Introduction}\label{sec:introduction}

\IEEEPARstart{T}{raffic} signal control systems (TSCSs) have been widely deployed for monitoring and controlling vehicular movements on the roads, by which the traffic flows can be effectively managed to ensure traffic safety~\cite{OliveiraIoT21}. 
A typical method is to control the color of traffic signals at the road intersections according to a fixed periodic schedule~\cite{OssamaIoT17}. However, this approach fails to incorporate real-time traffic conditions into the adjustment of traffic light status. Consequently, it may result in traffic congestion during peak traffic hours and energy wastage due to unnecessary control of traffic signal during off-peak periods. While manual traffic signal control can address this issue, its implementation at every road intersection in the real world is both costly and complex.

Intelligent TSCS is a viable solution to dynamically control the traffic signal cost-effectively\cpy{~\cite{ChenCIM22,DuiIoT24,GuAAAI24}}. Empowered by the state-of-the-art technologies including information sensing, data analysis, and optimization algorithms, the intelligent system is expected to automatically adjust the timing of traffic signal control according to real-time traffic conditions. Intelligent TSCSs offer significant advantages over static or manually controlled methods. These systems leverage real-time and historical data collected by monitoring sensors and cameras to make superior decision-making. Additionally, they have the ability to optimize the timing and cycle of traffic signal based on the current traffic flow, thereby improving the road condition.

Efficient control of traffic signals based on real-time status of intersections is crucial for achieving effective operation of TSCSs. The traditional centralized intelligent systems require to upload the collected traffic data to a remote central server. This process leads to an increase in data processing time and computational overhead, which can compromise the efficiency of the TSCSs~\cite{ArthursTITS22}. Edge intelligence is a promising technology that distributes the computational capability to the edge devices in the vicinity of the end users~\cite{XuIEEE21,ZhangIoT21}. Integrating edge intelligence to TSCSs enables the real-time decision-making, by which the latency associated with the data transmission and processing can be reduced. In this study, we explicitly illustrate the TSCSs based on edge intelligence, in which a multi-access edge computing (MEC) server is used to manage the traffic data collected from several road intersections (\textit{areas}). By leveraging the collaboration of multiple neighboring MEC servers for data acquisition, model training, and decision-making, the traffic signal status of all intersections across the entire area can be efficiently managed. 

Reinforcement learning algorithms have been widely studied to achieve effective traffic signal control\cpy{~\cite{WangTVT24,ZhouTITS24,JiangTITS24}}. Specifically, the problem of controlling traffic signals at multiple intersections can be formulated as a Markov decision process (MDP), wherein training can yield a policy for selecting the optimal action at each state~\cite{WangTCSS22}. Reinforcement learning algorithms can be applied to solve the MDP by learning the optimal decision on traffic signal control through value functions and policy optimization methods. Among these algorithms, soft actor-critic (SAC) is a potential candidate to extract the traffic information due to the inherent stochasticity of the policy with entropy regularization. State-of-the-art work has demonstrated the effectiveness of the SAC-based method in training optimally control decisions for traffic signal. The primary achievement includes attaining a consistently high traffic throughput while substantially reducing the average travel time for all vehicles~\cite{GeTITS22,MaoTITS23}.


In practice, road network is graph-structured, where intersections and roads can be represented as nodes and edges respectively. The recent advancement of graph neural network (GNN) has demonstrated the promising capability to automatically extract the information (traffic features) from adjacent intersections~\cite{NishiITSC18,RahmaniTITS23}. This aligns with the crucial role of traffic signals in influencing traffic conditions at intersections, which can in turn affect neighboring intersections. Furthermore, the current traffic conditions at an intersection can have an ripple effect on the traffic conditions at nearby intersections in subsequent moments. For this reason, by analyzing spatio-temporal information, TSCSs can dynamically adjust the duration of traffic lights to maximize traffic efficiency and reduce congestion. The recently proposed spatio-temporal GNN-based method has shown impressive success in reducing the average vehicular travel time~\cite{WangTMC22}. However, this method does not comprehensively take into consideration the influence that adjacent intersections exert on an individual intersection's traffic flow. Consequently, it might not be fully equipped to navigate the intricate and dynamic traffic conditions typical of real-world environments. The concept of a hypergraph, an extension of a traditional graph where an edge—referred to as a hyperedge—can connect more than two nodes, offers a promising alternative~\cite{AntelmiCSUR23,zhang2023learning}. This unique characteristic allows for the representation of more complex node correlations, such as those observed in real-world road traffic conditions. As a result, hypergraph is suitable for conducting higher-order network analysis with richer information.

By sharing traffic information among multiple MEC servers, each agent deployed in the road network can obtain the traffic conditions of other intersections. To facilitate the coordination among multiple agents for effective decision-making while enhancing their capability to process spatio-temporal information, we integrate hypergraph learning into the multi-agent SAC (MA-SAC). Specifically, the agents can dynamically construct spatio-temporal hyperedges and update embeddings by leveraging the heterogeneous properties of graphs. Therefore, the intelligent TSCS framework can dynamically adjust the phase sequence of traffic signals according to the traffic conditions at each intersection to promote seamless traffic flow across intersections. The main contributions of this work are summarized as below:
\begin{itemize}
    \item \textbf{Framework}: We propose a framework based on edge intelligence, which involves dividing the entire road network into multiple areas and deploying a MEC server in each area. This framework provides traffic information exchange between intersections and enables the training of multiple intelligent agents deployed at intersections.
    \item\textbf{Reinforcement Learning}: We adopt multi-agent SAC (MA-SAC), which is a stable off-policy algorithm based on the maximum entropy to the reinforcement learning framework. To enhance coordination in  controlling traffic conditions across multiple intersections, we implement an agent at each intersection. In addition to enabling agents interaction through the dynamical construction of hypergraph, we innovatively incorporate the loss generated by constructing hyperedges into the reinforcement learning loss, guiding the training of the learning process.
    
    \item \textbf{Hypergraph Learning}: Based on the proposed framework, we introduce hypergraph module to enable information interaction between intersections. In this module, we study the dynamic construction of spatial and temporal hyperedges to capture the spatio-temporal dependencies between multiple traffic signals. Our proposal offers greater flexibility compared to the traditional methods that solely consider the information of neighboring intersections in the surrounding area. To the best of the author's knowledge, this is the first work that introduces the concept of hypergraphs in the field of intelligent TSCSs.
    \item \textbf{Experiments}: We conduct extensive experiments on both synthetic and real-world traffic datasets. The experimental results demonstrate that our proposal can outperform state-of-the-art methods in various metrics including average travel time and throughput.

\end{itemize}

The remainder of the paper is organized as follows: Section~\ref{sec:relatedwork} summarizes the related work. Section~\ref{sec:problemstatement} introduces the problem formulation. 
Section~\ref{sec:algorithm} devises our proposed hypergraph-based deep reinforcement learning method. Section~\ref{sec:experiment} evaluates the performance of our proposal, and the paper is concluded with future work in Section~\ref{sec:conclusion}.


\section{Related Work}\label{sec:relatedwork}
This section discusses the state-of-the-art work in terms of traffic signal control based on reinforcement learning, integration of reinforcement learning with graph learning, and hypergraph learning.


\subsection{Traffic Signal Control Based on Reinforcement Learning}
The traditional transportation control method includes the timed-based control that uses a pre-determined plan for cycle length and phase time~\cite{Koonce2008TrafficST}, and that controls the signal that balances the queue length~\cite{Varaiya2013}. Recently proposed methods focus on utilizing reinforcement learning to realize self-adaptive traffic signal control in a single intersection~\cite{ChenAAAI2020}. 

Combining deep learning with reinforcement learning can further improve the learning performance of reinforcement learning in complex traffic scenarios~\cite{MaoITSM23}. For example, Wang \textit{et al.} proposed a double q-learning algorithm for traffic signal control~\cite{WangTOC21}. However, the value-based methods like DQN require complex operations to find the maximum reward value, making it difficult to handle high-dimensional action spaces. To address this issue, extensive researches have been conducted on policy-based reinforcement learning. Rizzo \textit{et al.} proposed a time critic policy gradient method to maximize the throughput while avoiding traffic congestion~\cite{RizzoKDD19}. Chu \textit{et al.} introduced a multi-agent deep reinforcement learning method based on advantage actor-critic (A2C)~\cite{ChuTITS20}. However, these methods only include the traffic information of the first-order neighbors of each agent, while those from higher-order are neglected. This results in the inaccurate estimation of traffic flow information within the road network, which in turn hinders the performance of traffic signal management.


\subsection{Integration of Reinforcement Learning and Graph Learning}
 Wei \textit{et al.} included the max-pressure metric in the reward function for traffic signal control~\cite{WeiHuaKDD19}. Neighbor RL is a multi-agent reinforcement learning method that directly connects the observation results from neighboring intersections to the state table~\cite{ArelIET}. However, both method only considers the information from the nearest intersections while ignoring that of the distant intersections. Specifically, the deep q-network (DQN) methods focus on using vector representations to capture the geometric features of the road network. Nevertheless, these algorithms are insufficient to cope with complex road network involving different topological structures.
 
 Graph neural networks (GNNs) can automatically extract features by considering the traffic features between distant roads by stacking multiple neural network layers\cite{RahmaniTITS23}. Nishi \textit{et al.} proposed a method based on reinforcement learning and graph learning, where GNNs are deployed to realize efficient signals control at multiple intersections~\cite{Nishi}. Specifically, they utilized the GNN method proposed in~\cite{Schlichtkrull} to extract the geometric features of the intersections. Wei \textit{et al.} proposed CoLight which utilizes graph attention networks to achieve traffic signal control. Specifically, to address conflicts in learning the influence of neighboring intersections on the target intersection, a parameter-sharing index-free learning approach that averages the influence across all adjacent intersections is employed~\cite{Colight}. 
 


\subsection{Hypergraph Learning}

Although the aforementioned standard graph-based methods are capable of forming pairwise relationships between two nodes, they fail to capture the multi-edge relationships among multiple nodes~\cite{zhang2019introducing}. For instance, in a real-world road network, an intersection is not only influenced by its immediate neighbors, but may also by a group of distant intersections. Standard graphs are typically represented by adjacency matrices to capture the relationships between nodes and edges. However, for the complex network systems like road network, individual nodes may have multiple features, resulting in multi-layer interactions among the nodes. The standard graph-based methods face challenges in effectively handling multimodal data. Additionally, the current traffic conditions at intersections are influenced  by the traffic conditions of neighboring roads in the previous time step. Therefore, it is essential to consider the spatio-temporal dependencies among different intersections when performing traffic signal control. As a prospective solution, the utilization of spatio-temporal graph neural networks (STGNNs) is anticipated to capture the spatial and temporal dependencies present within a graph. The implementation of STGNNs allows for a better understanding and utilization of the intricate relationships in multimodal data. Consequently, this enhances the capability of graph processing for multimodal data, enabling improved performance~\cite{wuIEEE}.

The emergence of hypergraph-based methods enables efficient handling of multi-modal data~\cite{GaoHGNN+}. In contrast to standard graphs where each edge can only connect with two nodes, hypergraphs introduce multiple hyperedges, which is capable of connecting any number of nodes. Therefore, hypergraphs can establish relationships beyond pairwise connections and have the potential to extract higher-order correlations from multiple related nodes. Hypergraph learning methods are gradually being utilized for clustering and classification problems. Arya \textit{et al.} employed geometric hypergraph learning to accomplish social information classification, yielding superior performance compared to pairwise standard graphs~\cite{AryaICMR18}. Wu \textit{et al.} proposed a hypergraph collaborative network that classifies both vertices and hyperedges~\cite{WuHypergraph}. Li \textit{et al.} employed geometric hypergraph learning and heterogeneous hypergraph clustering to achieve network motif clustering. The experimental results demonstrated that inhomogeneous partitioning offers significant performance improvements in various applications, including structure learning of rankings, subspace segmentation, and motif clustering~\cite{LiNIPS2017_a50abba8}. Wang \textit{et al.} constructed a hypergraph within the topological structure of subway systems and proposes a multi-layer spatio-temporal hypergraph neural network for subway traffic flow prediction. The effectiveness of the proposal is verified in terms of prediction accuracy~\cite{WangIEEETransactions}. Zhang \textit{et al.} proposed a hypergraph signal processing (HGSP) framework based on the tensor representation to generalize the standard graph signal processing (GSP) to tackle higher-order interactions~\cite{Zhang}.\setlength{\parskip}{0pt}

In summary, hypergraph combine multiple types of hyperedges to represent higher-order relationships, thus effectively addressing the association problem in multi-modal data. Given the presence of complex structural relationships in traffic network, this study introduces hypergraph to efficiently represent the traffic network. Within the hypergraph, spatial hyperedges and temporal hyperedges are dynamically generated based on the road network structure, resulting in a hypergraph with enhanced modeling capabilities. Subsequently, the intersection information is updated using multi-head attention, which facilitates dynamic spatio-temporal interactions among road intersections.

\subsection{Discussion}
\cpy{Recent researches like CoLight~\cite{Colight} have explored the application of reinforcement learning in traffic signal control by incorporating standard pairwise graph learning to model road network structures to enhance signal control across multiple intersections.  However, these approaches primarily focus on the traffic data from first-order neighbors at each intersection, which neglects the impact of more distant intersections and failing to consider the multi-edge relationships among various nodes. Furthermore, real-world datasets exhibit dynamic fluctuations in traffic flow that create higher-order correlations between intersections, which is critical for standard graph learning models to adequately capture. This limitation poses challenges in effectively managing complex road networks with diverse topological structures.}



\section{Problem Statement}\label{sec:problemstatement}
This section introduces the preliminaries on intelligent TSCS and presents the problem description with the problem objective. 



\subsection{Preliminaries}

\begin{figure}[!t]
\centering
\includegraphics[width=3in]{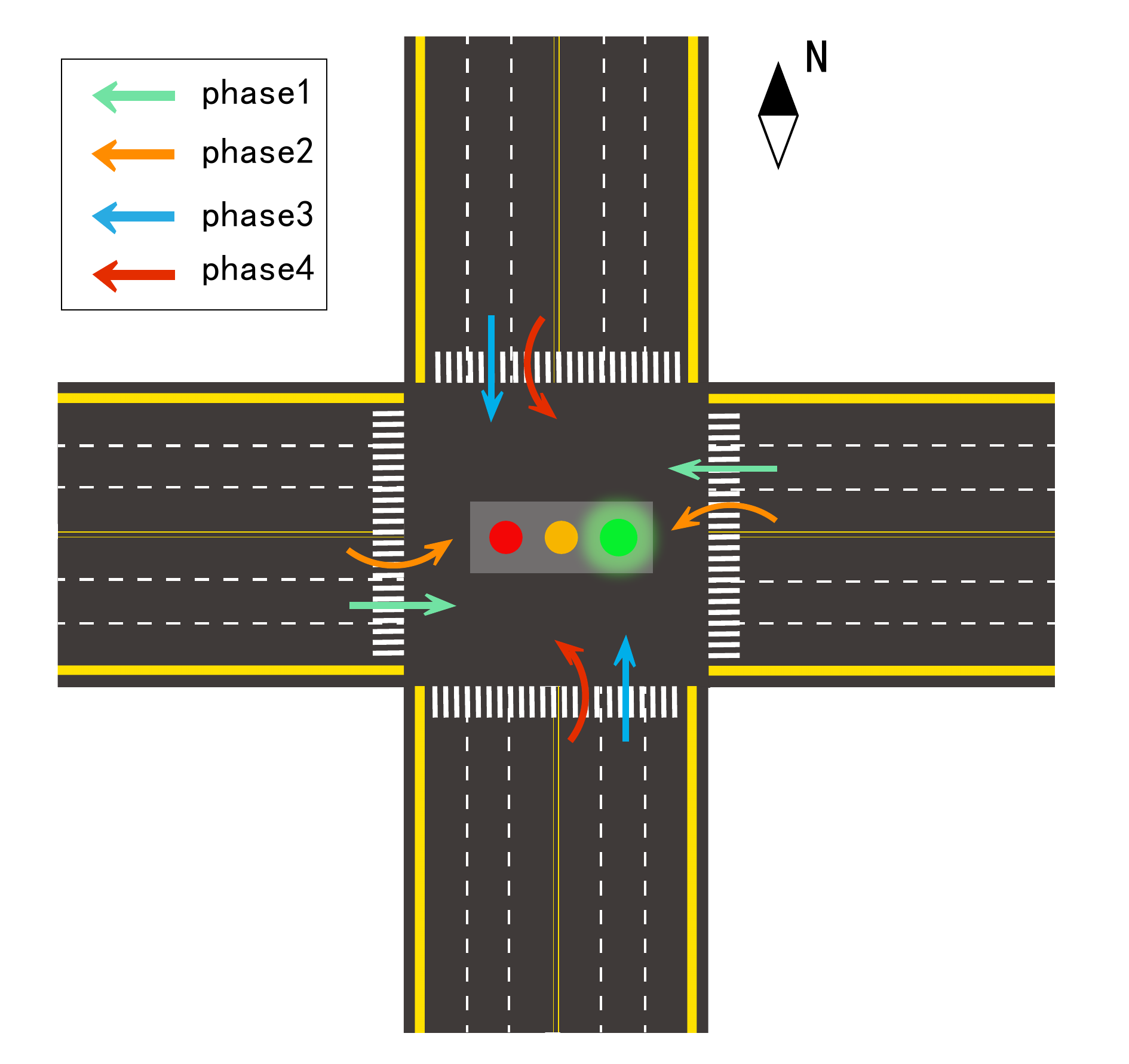}
\caption{A road intersection that includes four phases information.}
\label{fig:phase}
\end{figure}

In this paper, we divide the original road network into multiple circular areas, while a MEC server is deployed in each of them~\cite{WangTCSS22}. Each MEC server covers a certain area of the actual road network and collects traffic information around each intersection in the respective area. By sharing traffic information among the MEC servers, the information for various intersections in the entire road network can be acquired. Meanwhile, we assume a road traffic environment with three vehicle lanes and four movement directions, i.e., East (\textbf{E}), South (\textbf{S}), West (\textbf{W}), and North (\textbf{N}). Each intersection is equipped with a traffic light to control the traffic flows, i.e. Drive Through (\textbf{T}), Turn Left (\textbf{L}) and Turn Right (\textbf{R}), by switching among various phases. A phase is a combination of non-conflict vehicular movement signals. \figurename~\ref{fig:phase} illustrates a road intersection including four phases: Phase 1 (\textbf{ET} and \textbf{WT}), Phase2 (\textbf{EL} and \textbf{WL}), Phase3 (\textbf{ST} and \textbf{NT}), and Phase4 (\textbf{NL} and \textbf{SL}). When vehicles are making a right turn at the current intersection, they can be executed in all four phases mentioned above.

We use an agent to control the variation of phase for each intersection. Based on the information of the current intersection at time $t$, the agent selects one phase from the given four phases illustrated in \figurename~\ref{fig:phase}. The selected phase will be then implemented at the intersection for the time duration from $t$ to $t + \Delta t$. We set $\Delta t$ to 10 seconds to avoid excessive switching of actions by agents. The objective is to minimize the average travel time of all vehicles in the entire road network. After each change in the phase, a fixed interval (e.g., between 3 to 6 seconds) with a yellow signal will be conducted to regulate the traffic flow.

Regarding the network system, we first define the road network structure as a graph $G (V,E)$, where $V$ and $E$ denote a set of nodes and edges, respectively. $v_i$ $\in V$ and $e_i$ $\in E$ represent the observation information of $i$-th node and the $i$-th edge, which include the current phase of the traffic lights at the road intersection and the number of vehicles on each lane between the sender node and the receiver node. Agent $i$ is deployed at $i$-th road intersection, which is capable of observing the information of $i$-th node as well as that of all edges connected to this node. Based on the observed information ${o_i}$, agents autonomously select an appropriate phase stage.


Let $\mathcal{X}\!=\!\left\{\emph{O}^{1},...,\emph{O}^{T}\right\}\in\mathcal{R}^{T\times N\times d}$ denotes a spatial-temporal data containing $T$ timestamps, where $\emph{O}^{t}\!=\!\left\{\emph{o}^{t}_{1},...,\emph{o}^{t}_{N}\right\}\in \mathcal{R}^{N\times d}$ represents the signal of $N$ road intersections at the timestamp $t$, and the size of each road intersection's feature dimension is $d$. We define a hypergraph as $\mathcal{G} = \left\{\mathcal{V},\mathcal{E}\right\}$, where $\mathcal{V}$ is the set of node set, $\mathcal{E}=\{\emph{E}_{1},...,\emph{E}_{m}\}$ is a set of $m$ hyperedges in the hypergraph. $\emph{E}_{\alpha}$ is the hyperedge $\alpha$, which is an unordered set of nodes. Each hyperedge $e\in\mathcal{E}$ contains two or more nodes. $|e|$ is the size of hyperedge $e$. When each hyperedge $e\in\mathcal{E}$  satisfies $|e|=2$, the hypergraph will degenerate into the standard graph. In signal light control scenarios, using pairwise graphs to represent the relationships between intersections may lead to the loss of valuable information. Hypergraphs are more suitable for capturing the relevance of actual data with complex non-pairwise relationships.



\begin{figure*}[!t]
\centering
\includegraphics[scale=0.253]{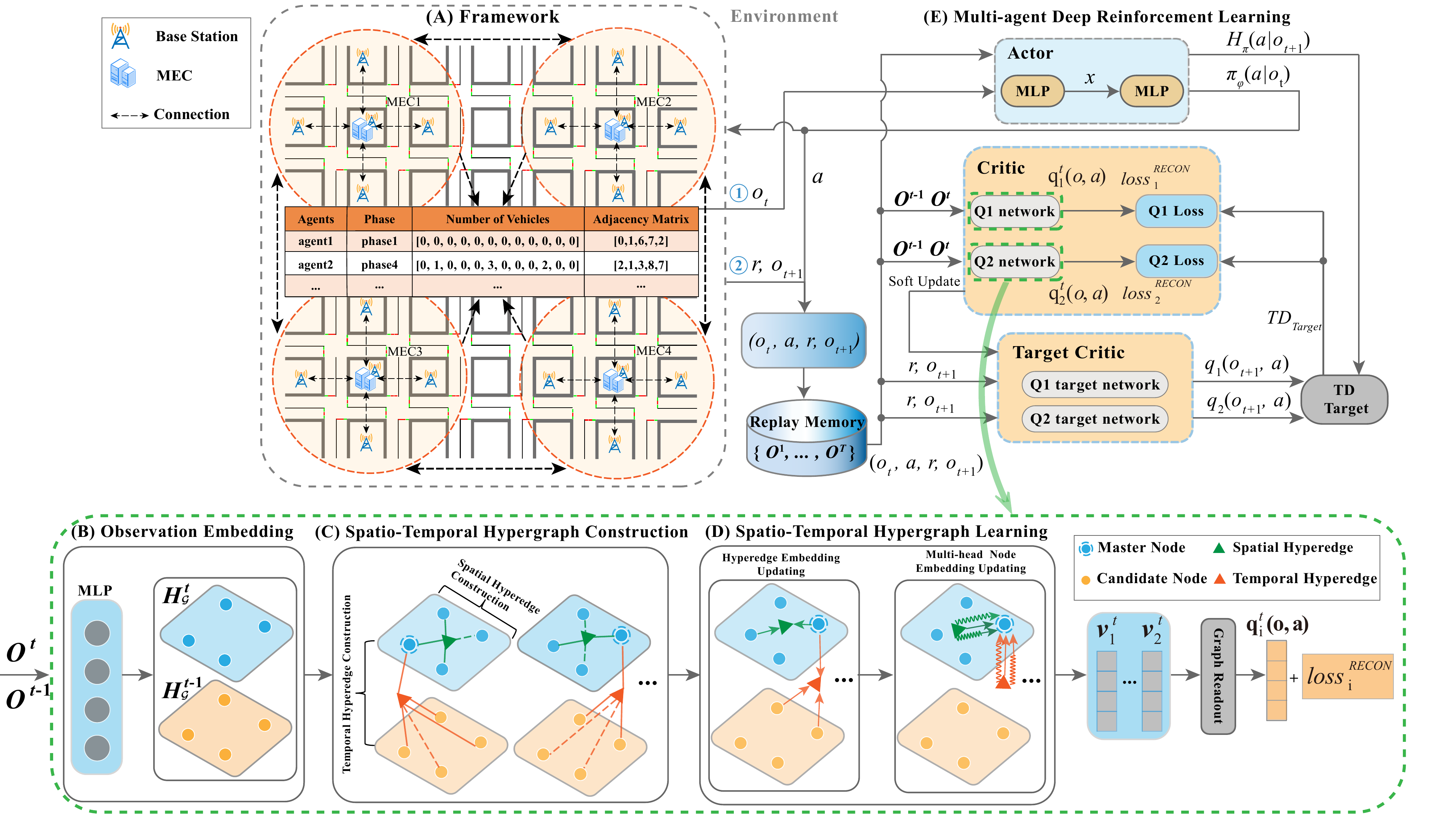}
\caption{An overview of our proposed framework.}
\label{fig:framework}
\end{figure*}

\subsection{Problem Description}
We define the problem of traffic signal control as a Markov decision process. The problem is characterized by the following major components:\begin{itemize}
    \item \textbf{State Space} $\mathcal{S}$: The environment state $s_t$ consists of the number of agents, the current phase of all agents, and the number of vehicles on each lane heading towards that agent at time $t$.
    \item \textbf{Observation Space} $\mathcal{O}$: At time $t$, each agent $i$ can only access a local observation $o_{i}^t$, which does not directly provide the complete system state, including all information in the traffic light adjacency graph. As illustrated in \figurename~\ref{fig:framework}, $o$ consists of agent information, its current phase, the number of vehicles on each lane connected with the intersection, and the adjacency matrix that to extract information about the agent and its neighbors.
     
    \item \textbf{Action} $\mathcal{A}$: At time $t$, each agent $i$ chooses an action $a_{i}^t$ (i.e., phase) for the next period of time. The actions taken by an intersection traffic signal control would impact its surrounding traffic flow, which in turn would influence its observations.

    \item \textbf{Transition probability} $\mathcal{P}$: $p(s_{t+1}|s_t,a_t)$ is the probability of transition from state $s_t$ to $s_{t+1}$ when all the agents take joint action $a_t$=\{$a_{i}^t\}_{i=1}^N \in \mathcal{A}$.

    \item \textbf{Reward}  $\mathcal{R}$: Agent $i$ obtains a reward $r_i^t$ by reward function $\mathcal{S}\times\mathcal{A}_1\times \cdots \times\mathcal{A}_N\rightarrow \mathcal{R}$. In this paper, the agent's objective is to minimize the travel time for all vehicles in the road network. \cpy{We define the reward for each intersection $n$ as:
\begin{equation}
    r_n^t = \text{-} \sum_l{u_{n,l}^t}
\end{equation}
where $u_{n,l}^t$ is the queue length on the approaching lane $l$ at time $t$. For each intersection, the waiting time for vehicles is influenced by the length of the queue in the lane. As the queue lengthens, the waiting time for vehicles increases, which results in a rise in overall travel time.}

    \item \textbf{Policy} $\pi$: At time $t$, agent $i$ chooses an action $a$ based on a certain policy $\mathcal{O} \times \mathcal{A} \rightarrow \pi$, aiming to maximize the total reward.  

    \item \textbf{Entropy} $\mathcal{H}$: Entropy is a measure of the randomness or uncertainty of a random variable. Specifically, $H(\pi(\cdot|s))$ represents the randomness or stochasticity of a policy $\pi$ in a given state $s$. By incorporating this entropy regularization term into the objective of reinforcement learning, we maximize the cumulative reward while promoting increased randomness in the policy $\pi$.
\end{itemize}

For the given road traffic network, each traffic signal agent $i$ at time $t$ makes action $a_{i}^t$ with reward $r_i^t$. The objective is to approximate the total reward $\sum_{\forall i}r_i^t$ by minimizing the loss. It is worth noting that, besides the loss defined in reinforcement learning algorithm, we additionally include the hypergraph reconstruction loss (See $\textit{loss}_i^\textit{RECON}$ in \figurename~\ref{fig:framework}) to enhance the training performance of the proposed framework. \cpy{We assume all agents are trained jointly while parameter sharing is not implemented due to the significant variation in the influence of each intersection's neighbors on the target intersection~\cite{Colight}}. More details will be described in Section~\ref{sec:algorithm}.
\section{Hypergraph-based Deep Reinforcement Learning Algorithm}\label{sec:algorithm}

This section introduces the details of our proposed hypergraph-based deep reinforcement learning for traffic signal control with edge intelligence.

\subsection{Framework}
\figurename~\ref{fig:framework} illustrates an overview of our proposed framework. In this framework, the road information at each intersection is collected from the MEC server. The role of the agent is to make an efficient control decision on the traffic signal phase for the respective intersection. Edge intelligence is used to enhance the data processing performance by sharing the necessary information among various MEC servers. \cpy{Edge intelligence enables computation close to the data source, thereby significantly reducing data transmission latency and bandwidth requirements. Moreover, edge intelligence allows each agent to rapidly process traffic data locally, facilitating real-time traffic information sharing between intersections and the training of multiple intelligent agents deployed at intersections. This collaborative mechanism can better address the complexities of traffic flow, leading to more efficient traffic signal control. }. 

The construction of hypergraph involves incorporating multiple surrounding agents to acquire the spatial and temporal structure information, i.e., $o_t$ includes  the state information of all intersections at time $t$. Meanwhile, the training process considers the historical state information as incorporating temporal dependency, i.e., ($o_t$, $o_{t-\Delta t}$) that covers the information at time $t$ and $t-\Delta t$. 

The proposed framework includes the following modules: 
\begin{itemize}
    \item \textit{Observation embedding}: It is an initialization module to obtain the current observation information $o_i^t$ from agent $i$.

    \item \textit{Hypergraph Learning}: The input of this module is the state information $\emph{O}^{t}$ and $\emph{O}^{t-1}$. The purpose of this module is to enhance the capabilities of both Critic Q1 and Q2 networks through the utilization of hypergraph learning. This module first creates spatial and temporal hyperedges separately to build node correlations, then it conducts hypergraph learning process to encode the attributes into the embedding space. One of the outputs of this module is the hypergraph reconstruction loss, which will be further utilized in the multi-agent reinforcement learning module to improve the training performance.
        
    \item \textit{Multi-agent deep reinforcement learning}: The input of this module is the agent $i$'s observation $o_i^t$ and reward $r$. A multi-agent soft actor-critic (MA-SAC) is introduced to obtain the best action $a_i^t$ that minimizes the predefined loss function. MA-SAC is composed of an actor network, two critic networks (Q1 and Q2 network), and two target-critic networks, which are all composed of fully connected neural networks.

\end{itemize}

\subsection{Observation Embedding}

Each MEC server collects and aggregates road information observed by agents within MEC's respective region, such as the count of vehicles in each lane and the current signal phase. We employ multiple fully connected neurons in the multi-layer perceptron (MLP) to achieve non-linear transformations and feature extraction on the input $d$-dimensional features. Rectified linear unit (ReLU) functions are applied to introduce non-linearity, resulting in an $l$-dimensional representation as:
\begin{equation}
    h_i^t = \text{Embed}(o_i^t) = \sigma(o_i^tW_e + b_e)
\end{equation}
where $o_i^t \in \mathbb{R}^{d}$ is the observation of agent $i$ at time $t$, and $d$ is the feature dimension of intersection $i$. $W_e \in \mathbb{R}^{d \times l}$ and $b_e \in \mathbb{R}^{l}$ are trainable weight matrix and bias vector. $\sigma(\cdot)$ is the ReLU function. The resulting hidden state $h_i^t \in \mathbb{R}^{l}$ represents the traffic information of the $i$-th intersection at time $t$. ${H}_{\mathcal{G}}^{t}=\{h_1^t ,..., h_N^t\}$ that represents the initial node embedding of the entire road network, will be served as the input of the following spatio-temporal hypergraph construction process.

\subsection{Spatio-temporal Hypergraph Construction}
First, we define the spatio-temporal hypergraph at timestamp $t$ as $\mathcal{G}^{t} = \left\{
\mathcal{V}^{t},\mathcal{E}^{t}\right\}$. To better explore the spatial correlation between intersections in road network, we aggregate the features of all intersections represented by $V^{t}\in R^{N\times d}$. By constructing spatial hyperedges, we can capture higher-order spatial correlations among multiple neighboring nodes compared to the standard graph. Then, to model the temporal dependencies between intersections along the time dimension, we utilize the nodes feature of one timestamp previous to capture the historical information \cite{zhang2023adaptive}. Thus, the node set $\mathcal{V}^{t}$ contains nodes (state information)  from two consecutive timestamps, $t-1$ and $t$,  which is defined as $\mathcal{V}^{t} = \left\{\emph{V}^{t},\emph{V}^{t-1}\right\}$. The node feature matrices for these timestamps are ${H}_{\mathcal{G}}^{t}$ and ${H}_{\mathcal{G}}^{t-1}$ respectively.


Owing to the dynamic nature of traffic flow and road conditions, employing initial hypergraph structures alone would overlook the dynamic modifications of such structures from adjusted feature embedding. As a result, some important implicit relationships may not be directly reflected in the inherent structural framework of hypergraphs.
Therefore, it is necessary to dynamically modify the hypergraph structure during the process of model optimization to adapt to the changes of traffic conditions in the road network. We propose a dynamic learning process that generates two types of hyperedges: \textit{spatial hyperedges} that capture heterogeneity between traffic light agents, encoding relations among different intersections in one timestamp, and \textit{temporal hyperedges} that scrutinize interactivity of historical traffic flow,  modeling the continuous interaction of road intersections in consecutive timestamps. During the dynamic hyperedges generation process, we introduce two kinds of nodes: master node and candidate node. A master node $\dot{v}\in\mathcal{V}$ acts as an anchor when generating the hyperedge $e\left(\dot{v}\right)\in\mathcal{E}$, combining with a set of candidate nodes $S\left(\dot{v}\right) = \left\{\hat{v}\right\}$ to collectively make up the hyperedge.

For each master node $\dot{v}^{t}_{i}\in\emph{V}^{t}$ in spatio-temporal hypergraph $\mathcal{G}^{t}$ at timestamp $t$, 
the spatial hyperedge $e_{spa}\left(\dot{v}_{i}^{t}\right)\in\mathcal{E}^{t}$ can be generated based on the reconstruction of the master node $\dot{v}^{t}_{i}$ and the spatial candidate node set $\Tilde{S}^{spa}\left(\dot{v}^{t}_{i}\right) = \left\{v|v\in\emph{V}^{t},v\neq\dot{v}^{t}_{i}\right\}$, which is denoted as:
    \begin{equation}
    {c}_{spa}\left(\dot{v}^{t}_{i}\right)= \|\emph{H}_{\mathcal{G}}^{t}\left(\dot{v}^{t}_{i}\right)\cdot\theta_{spa}-\emph{p}^{spa}_{\dot{v}^{t}_{i}}\cdot\emph{H}_{\mathcal{G}}^{t}\left(\Tilde{S}^{spa}\left(\dot{v}^{t}_{i}\right)\right)\|_{2}
    ~\label{eq:spa_cons}
\end{equation}
where ${c}_{spa}\left(\dot{v}^{t}_{i}\right)$ denotes the spatial reconstruction error. $\left\|\cdot\right\|_{2}$ denotes the $l2$ norm of the vector. $\emph{H}_{\mathcal{G}}^{t}\left(\dot{v}^{t}_{i}\right)$ and $\emph{H}_{\mathcal{G}}^{t}\left(\Tilde{S}^{spa}\left(\dot{v}^{t}_{i}\right)\right)$ are node feature matrices of the master node and the spatial candidate node set respectively. $\theta_{spa}$ is a specific trainable projection matrix when generating the spatial hyperedge $e_{spa}\left(\dot{v}_{i}^{t}\right)\in\mathcal{E}^{t}$.
$\emph{p}^{spa}_{\dot{v}^{t}_{i}}\in\mathcal{R}^{\left(N-1\right)}$ denotes the trainable reconstruction coefficient vector, 
with each element $\emph{p}^{spa}_{\dot{v}^{t}_{i}}\left(v\right)$ representing the learned reconstruction coefficient of each node $v\in\Tilde{S}^{spa}\left(\dot{v}^{t}_{i}\right)$ relative to the master node $\dot{v}^{t}_{i}$. According to $\emph{p}^{spa}_{\dot{v}^{t}_{i}}$, the nodes in the spatial candidate 
node set $\Tilde{S}^{spa}\left(\dot{v}^{t}_{i}\right)$ with reconstruction coefficient larger than the threshold $\zeta$  are selected to generate a spatial hyperedge of the master node $\dot{v}^{t}_{i}$ (See the connected through green solid lines in \figurename~\ref{fig:framework}). In contrast, unselected nodes with reconstructions coefficient values lower than $\zeta$ are connected by dotted lines. This facilitates the dynamic learning of the reconstruction coefficients between the master node and each other node within the road network. By comparing with threshold $\zeta$, we are able to dynamically filter nodes more appropriate for interactions, not merely acquiring the correlation coefficients of neighbor nodes to the master node. Consequently, our approach offers greater adaptability than traditional graph-based methods when it comes to handling the interplay of spatio-temporal information between intersections. This allows for a refined and contextual response to the ever-changing dynamics of traffic networks. $S^{spa}\left(\dot{v}^{t}_{i}\right)$ is denoted as: 
\begin{equation}
S^{spa}\left(\dot{v}^{t}_{i}\right) = \left\{v|v\in\Tilde{S}^{spa}\left(\dot{v}^{t}_{i}\right),\emph{p}^{spa}_{\dot{v}^{t}_{i}}\left(v\right)>\zeta\right\} 
\label{eq:zeta}
\end{equation}


Similarly, as shown in Fig.~\ref{fig:framework}, temporal candidate nodes are encircled from $\emph{V}^{t-1}$ to form the temporal hyperedge $e_{tem}\left(\dot{v}_{i}^{t}\right)$ with the master node based on the trainable reconstruction coefficient vector $\emph{p}^{tem}_{\dot{v}^{t}_{i}}\in\mathcal{R}^{ N}$. 
Overall, the loss of hyperedges generation in one timestamp is defined as:
\begin{equation}
\begin{split}
        \mathcal{L}_{recon}\!\! = \!\!\!\!\sum_{i=\left[1,...,N\right]}&\!\!\!\!\!\!\lambda\left(c_{spa}\!\left(\dot{v}^{t}_{i}\right)\!+\!c_{tem}\!\left(\dot{v}^{t}_{i}\right)\right)
        \!+\!\left(  \|\emph{p}^{spa}_{\dot{v}^{t}_{i}}\|_{1}+\|\emph{p}^{tem}_{\dot{v}^{t}_{i}}\|_{1}\right)\\&+{\gamma_2}\left(
    \|\emph{p}^{spa}_{\dot{v}^{t}_{i}}\|_{2}+\|\emph{p}^{tem}_{\dot{v}^{t}_{i}}\|_{2}\right)
    ~\label{eq:tem_cons}
\end{split}
\end{equation}
where $\left\|\cdot\right\|_{1}$ denotes the $l1$ norm of the vector and $c_{tem}$ denotes the reconstruction error of temporal hyperedges. $\lambda$ is the weight hyperparameter of the reconstruction error. $\gamma_2$ is the regularizing factor to balance $l1$ norm and $l2$ norm of the two types of reconstruction coefficient vectors.

\subsection{Spatio-temporal Hypergraph Learning}
\subsubsection{Hyperedge Embedding Updating}
Let $\emph{Re}$ represents the correlation between all nodes (within and not within the candidate node set $S^{spa}(\dot{v}^{t}_{i})$) in the road network and the  master node $\dot{v}^{t}_{i}$ at timestamps t, with entries are as below:
\begin{equation}
\emph{Re}\left(v,e\left(\dot{v}^{t}_{i}\right)\right) =\left\{ 
    \begin{matrix}
      1\text{,} & v = \dot{v}^{t}_{i} \\
    \emph{p}_{\dot{v}^{t}_{i}}\!\left(\!v\!\right)\!\text{,} & v \in S^{spa}\!\left(\!\dot{v}^{t}_{i}\!\right) \\
      0\text{,} & \text{otherwise}
    \end{matrix}\right.
\end{equation}
    The embedding of hyperedges is aggregated by node features as follows: 

\begin{equation}
    \emph{E}\left(e\left(\dot{v}^{t}_{i}\right)\right) = \frac{\sum_{v\in\mathcal{V}^{t}}\emph{Re}\left(v,e\left(\dot{v}^{t}_{i}\right)\right)\times\emph{H}_{\mathcal{G}}^{t}\left(v\right)}{\sum_{v\in\mathcal{V}^{t}}\emph{Re}\left(v,e\left(\dot{v}^{t}_{i}\right)\right)}
    \label{eq:edge_emb}
\end{equation}

\subsubsection{Multi-head Node Embedding Updating}
After implementing the update of hyperedges embeddings, we can utilize the information from the spatial hyperedge and temporal hyperedges to accomplish the update of the master node $\dot{v}^{t}_{i}$.
The hyperedges associated with node $\dot{v}^{t}_{i}$ is denoted as $\left\{e_{spa}\left(\dot{v}^{t}_{i}\right),e_{tem}\left(\dot{v}^{t}_{i}\right)\right\}$. 
We calculate multi-head attention between the master node and the two types of hyperedge, and subsequently employ the normalized attention as the weight for each hyperedge. The calculation process for the weights of the two types of hyperedge is similar, and we thus take calculating spatial hyperedge attention as an example:

\begin{equation}
att^{h}\left(\dot{v}^{t}_{i},e_{spa}\left(\dot{v}^{t}_{i}\right)\right) =  \frac{\emph{Q}^{h}\left(\dot{v}^{t}_{i}\right)\cdot\Theta_{spa}^{att}\cdot\emph{K}_{spa}^{h}\left(\dot{v}^{t}_{i}\right)^{T}}{\sqrt{d}}
\end{equation}
where
\begin{equation}
 \emph{Q}^{h}\left(\dot{v}^{t}_{i}\right)= \emph{H}_{\mathcal{G}}^{t}\left(\dot{v}^{t}_{i}\right)\cdot Q\text{-}Lin^{h} 
\end{equation}
\begin{equation}
    \emph{K}_{spa}^{h}\left(\dot{v}^{t}_{i}\right) = \emph{E}\left(e_{spa}\left(\dot{v}^{t}_{i}\right)\right)\cdot K\text{-}Lin^{h}_{spa}
\end{equation}

First, for the $h$-th attention head $att^{h}\left(\dot{v}^{t}_{i},e_{spa}\left(\dot{v}^{t}_{i}\right)\right)$, we apply a linear transformation matrix $Q\text{-}Lin^{h}\in\mathcal{R}^{d\times\frac{d}{K}}$ to project the feature information of the master node $\dot{v}^{t}_{i}$ into the $h$-th query vector $\emph{Q}^{h}\left(\dot{v}^{t}_{i}\right)$. The value of $K$ is the number of attention heads. 
Additionally, We project the spatial hyperedge $e_{spa}\left(\dot{v}^{t}_{i}\right)$ into the $h$-th key vector $\emph{K}_{spa}^{h}\left(\dot{v}^{t}_{i}\right)$ on the same dimension. Next, we apply a trainable weight matrix $\Theta_{spa}^{att}\in\mathcal{R}^{\frac{d}{K}\times\frac{d}{K}}$ to obtain $h$-th spatial attention, and $\sqrt{d}$ acts as a scaling factor. The $h$-th temporal attention $att^{h}\left(\dot{v}^{t}_{i},e_{tem}\left(\dot{v}^{t}_{i}\right)\right)$ is calculated in a similar manner. Finally, the weight of the spatial hyperedge $w^{h}_{spa}\left(\dot{v}^{t}_{i}\right)$ and the temporal hyperedge $w^{h}_{tem}\left(\dot{v}^{t}_{i}\right)$ are calculated by $softmax$ normalization to $att^{h}\left(\dot{v}^{t}_{i},e_{tem}\left(\dot{v}^{t}_{i}\right)\right)$ and $att^{h}\left(\dot{v}^{t}_{i},e_{spa}\left(\dot{v}^{t}_{i}\right)\right)$. The attentive aggregation of different heads among hyperedges for updating node embedding of $\dot{v}_{i}^{t}$ (referred to as wavy lines in \figurename~\ref{fig:framework} is denoted as :
\begin{equation}
\begin{split}    
\emph{Q}_{\emph{V}^{t}}\left(\dot{v}^{t}_{i}\right) = MLP\Bigl(\mathop{\concat}\limits_{h\in\left[1,K\right]}&(w^{h}_{spa}\left(\dot{v}^{t}_{i}\right)\times \emph{K}_{spa}^{h}\left(\dot{v}^{t}_{i}\right)\\&+w^{h}_{tem}\left(\dot{v}^{t}_{i}\right)\times \emph{K}_{tem}^{h}\left(\dot{v}^{t}_{i}\right))\Bigl)
\end{split}
~\label{eq:att}
\end{equation}
where $\concat$ represents concatenation. We first aggregate two types of hyperedges associated with the master node on the $h$-th attention head, and then concatenate the results from all $K$ heads. Subsequently, the node embedding of $\dot{v}^{t}_{i}$ is updated by a shallow MLP. We average the node embedding of all nodes $\emph{Q}_{\emph{V}^{t}}$ in road network at timestamp $t$ to read-out the graph representation of $\mathcal{G}^{t}$, which is denoted as $\emph{Q}_{\mathcal{G}^{t}}\in\mathcal{R}^{d}$.


To realize the procedure in an end-to-end fashion, the loss function used in the training process is denoted as :
\begin{equation}
    \mathcal{L}_\textit{HG} = \beta\mathcal{L}_{recon}+\left(1-\beta\right)MSE\left(MLP\left(\emph{Q}_{\mathcal{G}^{t}}\right),\emph{y}^{t}\right)
    ~\label{eq:HG_loss}
\end{equation}
where \textit{MSE} is the mean squared error function. $\beta$ is a weight hyperparameter to balance the impact of reconstruction loss and Q-value loss. $y^t$ is the target Q-value (referred to as $\textit{TD}_\textit{target}$ in \figurename~\ref{fig:framework}) 

 
\subsection{Multi-agent Deep Reinforcement Learning}


We introduce MA-SAC based on the entropy-regularized reinforcement learning for traffic signal control. In entropy-regularized reinforcement learning, the agent gets a reward $r$ at each time $t$ proportional to the entropy of the policy at that time. The objective of MA-SAC is to not only optimize the cumulative expected rewards, but also maximize the expected entropy of the policy as:
\begin{equation}
    \pi^{*} = \arg\max_{\pi}\mathbb{E}_{\pi}\left(\sum_{\forall t}r(o_t,a_t)+\alpha H(\pi_t(\cdot|o_t))\right)
\end{equation}
where $\alpha$ is temperature parameter that determines the relative importance of the entropy term versus the reward~\cite{Petros2019SoftAF}. $H(\pi_t(\cdot|o_t))$ is the entropy calculated by:
\begin{equation}
    H(\pi_t(\cdot|o_t))=-\sum_{a}\pi_t(a_t|o_t)\log \left(\pi_t(a_t|o_t)\right)
\end{equation}

To assist the automatic process of tuning the entropy regularization, the expected entropy can be limited to be no less than an objective value $\mathcal{H}_0$, and thus, the objective can be reformulated as a constrained optimization problem as:

\begin{equation}
    \pi^{*} = \max_{\pi}\mathbb{E}_{\pi}\left(\sum_{\forall t}r(o_t,a_t)\right)
\end{equation}
s.t.
\begin{equation}
    \mathbb{E}_{(o_t,a_t)\sim \rho_\pi}\left(-\log (\pi_t(a_t|o_t))\right)\geq \mathcal{H}_0
\end{equation}

During the process of adjusting the entropy regularization coefficient for each agent, the temperature parameter $\alpha$ is utilized to improve the training performance of actor network. The value of $\alpha$ increases when the entropy obtained from the actor network is lower than the target value $\mathcal{H}_0$. In this way, it increases the importance of the policy entropy term in the actor network's loss function. Conversely, decreasing the value of $\alpha$ will reduce the importance of the policy entropy term and allow for a greater focus on the reward value improvement. Therefore, the choice of the entropy regularization coefficient 
$\alpha$ is crucial. By automatically adjusting the entropy regularization term, we derive the loss function for $\alpha$ using simple mathematical techniques so that we do not need to set it as a hyperparameter. The loss function of $\alpha$ is defined as:
    \begin{equation}
        L(\alpha)=\mathbb{E}_{\pi}[-\alpha\log\pi(a_{t}|o_{t})-\alpha\mathcal{H}_0]
        \label{eq:alpha_loss}
    \end{equation}

MA-SAC adopts an actor-critic architecture with policy and value networks, which can reuse the collected data and entropy maximization to achieve effective exploration. As illustrated in the upper-right part of \figurename~\ref{fig:framework}, MA-SAC consists of six networks: an actor network and four critic networks (Q1 network, Q2 network, Q1 target network, and Q2 target network) and an $\alpha$ network. The summary of these networks is as below: \begin{itemize}
    
    \item \textbf{Actor network}: The actor network is responsible for exploring and selecting actions based on the learned policy, which influences the behavior of the agent in the multi-agent environment. The input includes the batch size, the number of intersections and state. Based on the probability that each action to be executed, the training results aim to maximize the overall state value $V^{\pi}(o)$ calculated by:
    \begin{equation}
        V^{\pi}(o) =  \mathbb{E}_{\pi}\left(Q^{\pi}(o_t,a_t)+\alpha H(\pi_t(\cdot|o_t))\right)
    \end{equation}
    where $Q^{\pi}(o_t,a_t)$ is the Q-function expressed by:
    \begin{equation}
        Q^{\pi}(o_t,a_t) = \mathbb{E}_{\pi}\left( r(o_t,a_t,o_{t+1}) + \gamma V^{\pi}(o_{t+1})\right)
    \end{equation}
    The actor network outputs the probability of the available actions as $\pi_{\phi}(a_t|o_t)$ and its respective entropy value as $H_{\pi}(a_t|o_t)$.
    
    \item \textbf{Critic network}: The critic network is designed to calculate the Q value to evaluate the action. As illustrated in \figurename~\ref{fig:framework}, the critical Q-network represents the estimation of the action value while the target critic network indicates the estimation of the state value. To stabilize the training process, the update frequency of target critic networks is less than that of the critic Q-network. Different from the actor networks, the output of the critic network is the value of Q function as $q_1(o_{t+1},a_t)$ and $q_2(o_{t+1},a_t)$.  
    
\end{itemize}

Algorithm~\ref{alg:SAC} shows the details of our developed MA-SAC algorithm. This algorithm mainly consists of two phases: experience gathering and network training. In experience gathering phase, the experience replay mechanism is incorporated to mitigate the correlation among data samples. Each agent $i$ performs the actions $a$ generated in each episode, and then stores the tuples $(o_t,a_t,r_t,o_{t+1})$ into the replay buffer (line \ref{line:select} to \ref{line:store} in Algorithm~\ref{alg:SAC}).

The training stage begins once the data in the replay buffer reaches the given threshold $\textit{Thres}_\textit{size}$. During each step, some data will be randomly chosen from the replay buffer to update the parameters of both the actor networks and critic networks. The actor network updates the target by:
\begin{equation}
\begin{split}
    \mathcal{J}  =&  \mathbb{E}_{o_t,a_t}(\alpha \log(\pi_i(a_i|o_i)) \\&-Q_i^{\pi}(o_t,a_1,...,a_N)|_{a_i=\pi_i(o_i)}), i=1,2
    \end{split}
    \label{eq:actor_loss}
\end{equation}

The target for Q functions is expressed by:
\begin{equation}
      \begin{split}
       y_i &=  r_i + \\&\gamma \mathbb{E}\left(\min_{i=1,2}{Q_{\phi_{\text{targ},i}}{(o_{t+1},\tilde{a}_{t+1})}} - \alpha \log \pi_{\theta}(\tilde{a}_{t+1}| o_{t+1})\right),\\& \tilde{a}_{t+1} \sim \pi_{\theta}(\cdot| o_{t+1})\;
       \end{split}
       \label{eq:target}
       \end{equation}

Based on the obtained targets, the critic networks are updated by minimizing the loss function calculated by:
\begin{equation}
    \mathcal{L} =
    (1-\beta)
    \mathbb{E}_{o_t,a_t,r_t,o_{t+1}}\left(Q_i^{\pi}(o_t,a_1,...,a_N)-y_i\right)^2+\beta \mathcal{L}_{recon} 
    \label{eq:loss}
\end{equation}
where $\beta$ is a weight hyperparameter to trade off the effects of temporal difference (TD) target and the loss of hyperedges generation (See Equation~\ref{eq:tem_cons}) to enhance the training capability of Q1 and Q2 networks in hypergraph learning module.

In order to ensure the stability of training, the parameters of both actor networks and critic networks are copied to the corresponding target networks using the soft update method. This process is illustrated in line~\ref{line:targetnw} of Algorithm~\ref{alg:SAC}, where $\rho$ is the update ratio.

\begin{algorithm}[ht!]
\SetKwInput{Input}{Input}
\SetKwInput{Output}{Output}
\SetKwFor{For}{for}{}{end}
\SetKwFor{When}{when}{}{end}
\SetKwFor{Foreach}{for each}{}{end}
\SetKw{GoTo}{go to}
\SetKw{Return}{return}
\SetKw{Repeat}{Repeat}
\SetKw{Until}{Until}{}
\caption{Multi-agent Soft Actor-Critic (MA-SAC)}\label{alg:SAC}

Initial policy parameters $\theta$, Q-function parameters $\phi_{1}$, $\phi_{2}$, empty replay buffer $\mathcal{D}$\;
Set target parameters equal to main parameters $\phi_{\textit{targ}.1}\leftarrow \phi_{1}$, $\phi_{\textit{targ}.2}\leftarrow \phi_{2}$\;
\Repeat
Observe state $s$ and select action $a \sim \pi_\theta(\cdot |o_t)$\;\label{line:select}
Store ($o_t,a_t,r_t,o_{t+1}$) in replay buffer \textbf{$\mathcal D$}\;\label{line:store}
   \For{each training step }{
       Sample a random mini-batch of transitions, $\mathcal{B}= \{(o_t,a_t,r_t,o_{t+1}) \}$ from \textbf{$\mathcal D$}\;
       Compute targets for the Q functions by Equation~\ref{eq:target}\;
       Update critic networks by Equation~\ref{eq:loss}\;
       Update actor networks by Equation~\ref{eq:actor_loss}\;
       Update temperature by Equation~\ref{eq:alpha_loss}\;
       Update target networks with $\phi_{targ,1}\leftarrow \rho\theta_{\textit{targ},1} + (1-\rho)\phi_1$,  ~\label{line:targetnw} 
       $\phi_{targ,2}\leftarrow \rho\theta_{\textit{targ},2} + (1-\rho)\phi_2$\;
   }
\Until{convergence}

\end{algorithm}


\section{Experiments}\label{sec:experiment}
This section presents a detailed description of the extensive experiments conducted to analyze the performance of our proposed method and validate its effectiveness in traffic signal control.

\subsection{Experimental Settings}
 We conduct the experiments on CityFlow~\cite{CityFlow}, which is an open-source traffic simulator that supports city-wide large-scale traffic signal controller and flexible definitions for road network and traffic flow. After feeding the given traffic datasets into the simulator, a vehicle moves towards its destination according to the environment setting. This simulator provides the road state based on the given traffic signal control method and simulates the behavior of individual vehicles, providing the detail of traffic evolution in a road network. 

We introduce four synthetic traffic datasets ($\textit{Unidirect}_\textit{6$\times$6}$, $\textit{Bidirect}_\textit{6$\times$6}$, $\textit{Unidirect}_\textit{10$\times$10}$, and $\textit{Bidirect}_\textit{10$\times$10}$ ) and two real-world traffic datasets ($D_\textit{Hangzhou}$ and $D_\textit{Jinan}$) to validate the robustness of our proposed method. The details of them are as below: \begin{itemize}
    \item Synthetic dataset: $\textit{Unidirect}$ and $\textit{Bidirect}$ indicate uni- and bi-directional traffic, respectively. 6$\times$6 and 10$\times$10 represent the size of network, i.e., a 6$\times$6 or a 10$\times$10 grid network. For these datasets, 
    each intersection has four directions as East (\textbf{E}), South (\textbf{S}), West (\textbf{W}), and North (\textbf{N}). Each direction is with 3 lanes (300 meters in length and 3 meters in width). In bi-directional traffic, vehicles arrive uniformly with 300 vehicles/lane/hour in W$\leftrightarrow$E direction and 90 vehicles/lane/hour in S$\leftrightarrow$N direction. Only W$\rightarrow$E and N$\rightarrow$S directional flows travel in uni-directional traffic.
    \item Real-world dataset:  $D_\textit{Hangzhou}$ and $D_\textit{Jinan}$ are publicly available real-world traffic data for Hangzhou City and Jinan City, respectively. The traffic flows are processed from multiple sources. The number of intersections is 16, and 12 for these two datasets respectively. More details can be found in \cite{Colight}.

\end{itemize} 


The main parameter settings of our proposed methods are summarized in Table~\ref{tab:params}.

\begin{table}[tb]
 \caption{Main parameter settings.}
 \label{tab:params}
 \centering
 \begin{tabular}{l|l}
   \hline
   Parameter  & Value \\
   \hline 
   Batch size & 20\\
   Episodes & 50\\
   
   Target entropy & -0.5\\
   Learning rate & 0.0001 (actor), 0.01 (critic), and 0.001($\alpha$)\\
   Replay memory ($\textit{Thres}_\textit{size}$) & 1000\\
   Optimizer & Adam\\
   Head number (K) & 1\\
   Discount rate ($\gamma$) & 0.98\\
   Reconstruction error ($\lambda$) & 0.001\\
   Regularizing factor ($\gamma_2$) & 0.2\\
   Update ratio ($\rho$) & 0.005 \\

   \hline
  \end{tabular}
\end{table}

\subsection{Baseline Methods}
We verify the performance of our proposed hypergraph-based deep reinforcement learning method, referred to as  \textbf{\algname}, with 6 comparative methods. These methods include the traditional transportation methods,  reinforcement learning methods, and some that incorporate graph learning. \cpy{To ensure a fair comparison, we utilize the optimal parameters configurations for various datasets as specified in the original open sourcecodes of the baseline methods.} The details of the baseline methods are described as below: 
\begin{itemize}
    \item \textbf{Fixed-time}~\cite{Koonce2008TrafficST}: Employing a pre-defined plan for cycle length and phase time in traffic light control. It is widely applied when the traffic flow is steady.

    \item \textbf{MaxPressure}~\cite{Varaiya2013}: Implementing traffic signal control by relieving the vehicles on the lane with maximum pressure (a pre-defined metric about upstream and downstream queue length). 
\end{itemize}
\begin{itemize}
    \item \textbf{PressLight}~\cite{WeiHuaKDD19}: Utilizing the concept of MaxPressure and deep reinforcement learning to effectively optimize the pressure at each intersection. This method is designed to solve the multi-intersection signal control problems. \cpy{The batch size is 20, episodes is 100, learning rate is 0,001, replay memory is 3000, discount rate is 0.8, and Adam is used as optimizer.}

    \item \textbf{MPLight}~\cite{ChenAAAI2020}: This method utilizes the concept of pressure to achieve signal coordination at the regional level in a reinforcement learning-based approach, while also employing a network structure specifically designed to handle unbalanced traffic flow. \cpy{The key parameters are configured to match those used in PressLight.}

    

    \item \textbf{CoLight}~\cite{Colight}: A graph attention network is introduced in the reinforcement learning setting of multi-intersection traffic signal control to enhance coordination and decision-making. \cpy{The replay memory is 1000, while RMSProp is used as optimizer. Other main parameters are set as the same as those in Presslight.}
    

    
    \item \textbf{GCN-SAC}: We design this method for conducting ablation experiments to validate the performance between standard graphs and hypergraphs. In the reinforcement learning module, we set GCN-SAC to be the identical as HG-DRL, while in GCN-SAC, graph attention network (GAT) is used to facilitate information interaction between adjacent nodes. \cpy{The key parameters are configured to match those used in HG-DRL.}
    
\end{itemize}

\begin{table*}[ht]
 \caption{Performance of average travel time (in seconds) for various methods.}
 \label{tab:traveltime}
 \centering
\begin{tabular}{l|cccccc}
   \hline
   Methods & $\textit{Unidirect}_\textit{6$\times$6}$ & $\textit{Bidirect}_\textit{6$\times$6}$ & $\textit{Unidirect}_\textit{10$\times$10}$ & $\textit{Bidirect}_\textit{10$\times$10}$  & $D_\textit{Hangzhou}$ & $D_\textit{Jinan}$ \\
    \hline
    Fixed-time   & 210.94& 210.94 &345.82&345.82 & 718.29 & 814.11\\
    MaxPressure    &186.56 & 195.49 &297.18&322.71  & 407.17 & 343.90 \\\hline
    PressLight  &365.58 &256.89  &313.23&349.07 &421.18  &354.93 \\
    MPLight   &214.69 &238.27  &297.38&311.05  & 324.67 & 334.01     \\
    CoLight   &174.73 & 173.88 &347.87&352.75  &  301.75 &  326.75   \\
    GCN-SAC &173.37&172.31& 297.75  &306.16&336.15&332.36          \\\hline
    \textbf{\algname}&\textbf{168.85 \cpy{$\pm$ 2.70}}&\textbf{167.77 \cpy{$\pm$ 1.01}}&\textbf{286.87 \cpy{$\pm$ 2.42}}&\textbf{285.85 \cpy{$\pm$ 1.75}}&\textbf{295.71 \cpy{$\pm$ 2.04}}&\textbf{278.22  \cpy{$\pm$ 3.46}}\\\hline

\end{tabular}
\end{table*}

\begin{table*}[ht]
 \caption{Performance of throughput for various methods.}
 \label{tab:throughput}
 \centering

\begin{tabular}{l|cccccc}
   \hline
   Methods & $\textit{Unidirect}_\textit{6$\times$6}$ & $\textit{Bidirect}_\textit{6$\times$6}$ & $\textit{Unidirect}_\textit{10$\times$10}$ & $\textit{Bidirect}_\textit{10$\times$10}$  & $D_\textit{Hangzhou}$ & $D_\textit{Jinan}$ \\
    \hline
    Fixed-time   &2190 &4380 &3480&6960&1989&3475\\
    MaxPressure   &2214 &4408 &3542 &7036&2664&5613\\\hline
    PressLight    &2295&4627&3859&7736&2802&6043\\
    MPLight  &2313  &4640 &3866&7736&2923&5927    \\
    CoLight  &2327 &4652 &3865 &7743 &2929 &6061  \\
    GCN-SAC &2324 &4652&3870&7744&2915&6069     \\\hline
    \textbf{\algname}&\textbf{2327 \cpy{$\pm$ 3}}&\textbf{4652 \cpy{$\pm$ 2} }&\textbf{3870 \cpy{$\pm$ 6}  }&\textbf{7755 \cpy{$\pm$ 4} }&\textbf{2932 \cpy{$\pm$ 4} }&\textbf{6149 \cpy{$\pm$ 5} }\\\hline

\end{tabular}
\end{table*}

\subsection{Evaluation Metric}
\subsubsection{Travel Time} It is defined as the average travel time (\textbf{ATT}) of all vehicles spend
between entering and leaving the area. This value is affected by the factors such as red/yellow 
 traffic signals and traffic congestion. Travel time is a widely adopted criterion to measure the performance of traffic signal control~\cite{MeiML2023}. 
\subsubsection{Throughput} It indicates the number of vehicles that have finished their trips until the current simulation step. A larger throughput value indicates the better performance in traffic signal control. 

It is worth noting that ATT represents the average travel time of all vehicles across the entire road network within a given time period. In cases where vehicles enter the network but do not exit it, a simulated end time is assigned as their exit time from the current network. Consequently, these vehicles, which enter the network later and do not exit, play a role in reducing the overall ATT for all vehicles. Therefore, throughput also serves as a crucial evaluation metric, with higher values indicating that a greater number of vehicles have successfully completed their trips within the specific time period. This signifies improved control performance of a given method.

\subsection{Performance Evaluation}

 Tables~\ref{tab:traveltime} and \ref{tab:throughput} summarize the performance of our proposed \algname against the classic transportation methods as well as state-of-the-art reinforcement learning methods in both synthetic and real-world datasets. The results demonstrate that our proposed \algname outperforms other comparative methods in all datasets, achieving the minimum ATT for all vehicles entering the road network while maximizing throughput. These findings indicate that our proposal exhibits superior traffic signal control strategies. 

 In details, \algname reduces ATT by 2.60$\%$ and 2.63$\%$ compared to the second-best method GCN-SAC in $\textit{Unidirect}_\textit{6$\times$6}$ and $\textit{Bidirect}_\textit{6$\times$6}$ respectively.
When the road network scale expands to  10$\times$10, the advantage of our proposal against the second-best method grows to 3.46$\%$ and 6.63$\%$ for uni-direction and bi-direction topology respectively. This is because our introduced hypergraph can capture abundant higher-order temporal and spatial correlations between multiple intersections in large-scale road network. Meanwhile, in real-world dataset $D_\textit{Jinan}$, \algname can reduce ATT by 14.85$\%$ compared to the second-best method Colight. This is due to the fact that the driving paths of vehicles in real datasets are more complex than the synthetic datasets, and our proposed method can dynamically process key information from multiple upstream and downstream intersections based on historical road conditions. 

\begin{figure}[ht]
	\centering
        \begin{minipage}[b]{0.493\columnwidth}
		\centering
		
		\includegraphics[width=\columnwidth]{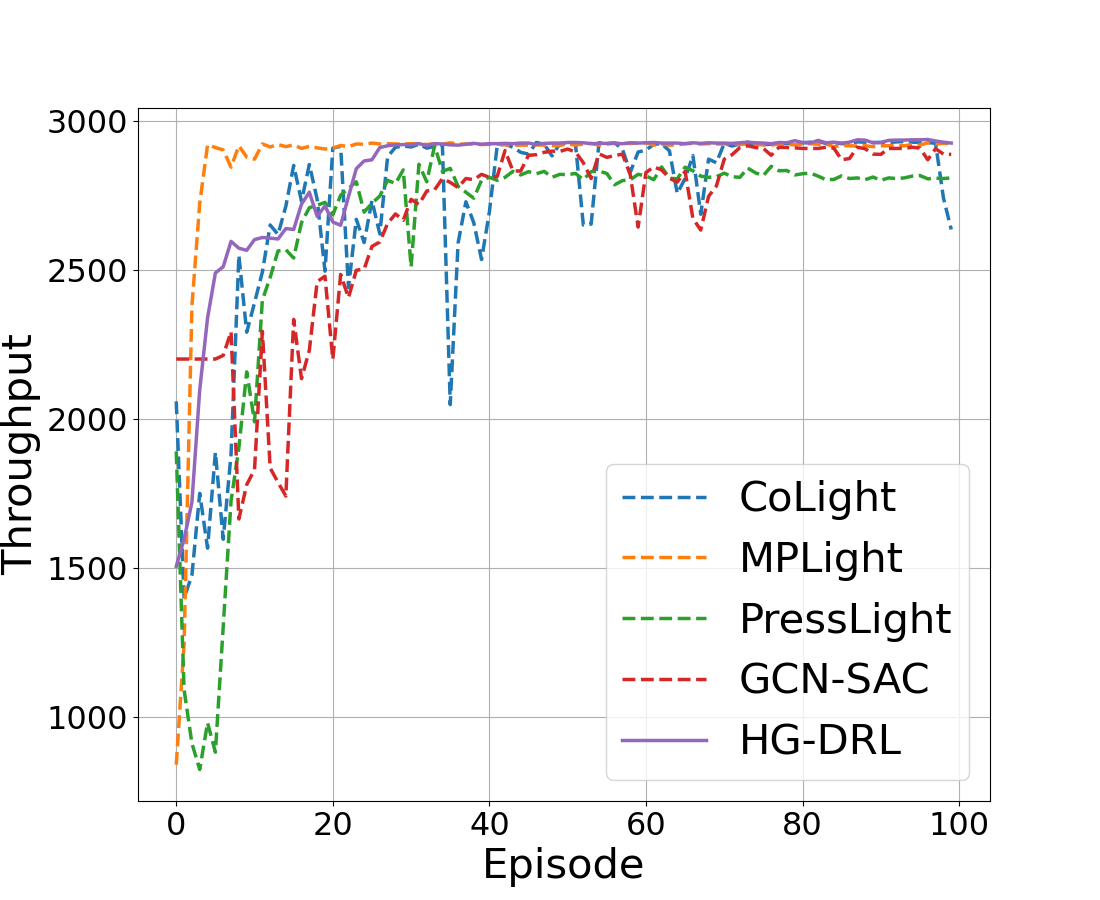}
		\subcaption{$D_\textit{Hangzhou}$}\label{fig:thr_hangzhou}
	\end{minipage}
	\begin{minipage}[b]{0.493\columnwidth}
		\centering
		\includegraphics[width=\columnwidth]{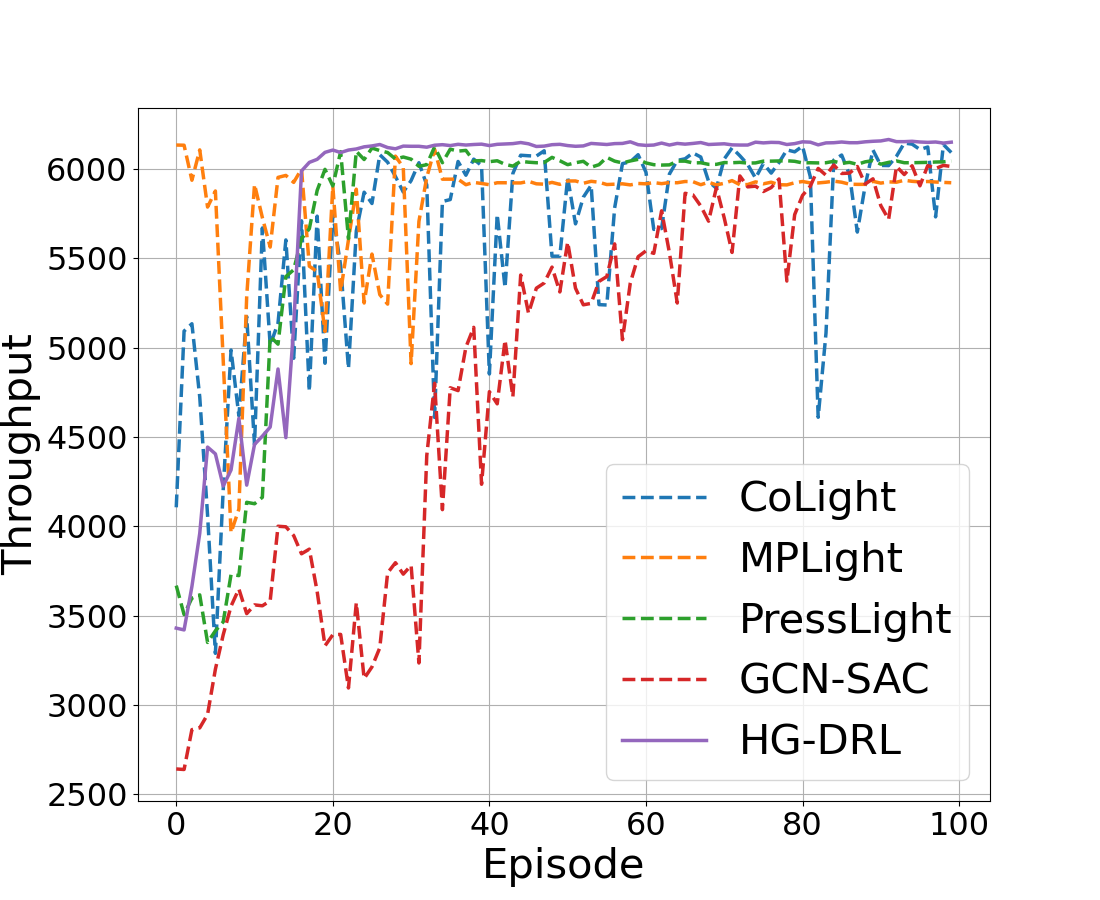}
		\subcaption{$D_\textit{Jinan}$}\label{fig: thr_jinan}
	\end{minipage}
	\caption{\cpy{Throughput evolution for different methods.}}	
\label{fig:thr}
\end{figure}

\newinfo{In terms of the learning curves for different methods, our proposed method HG-DRL demonstrates enhanced post-convergence stability compared to other baseline methods. This stability arises from HG-DRL's ability to dynamically construct hyperedges, enabling efficient capture and representation of complex spatio-temporal correlations inherent in traffic signal control systems. For example, CoLight demonstrates notable fluctuations in result values across a broad range of episodes, whereas our proposed method maintains consistent stability within the same range. This further substantiates the capability of HG-DRL to sustain stability across a wide range while achieving optimal performance. These findings align with the trends observed in the throughput evolution, as shown in \figurename~\ref{fig:thr}. }

\begin{figure}[ht]
	\centering
        \begin{minipage}[b]{0.493\columnwidth}
		\centering
		
		\includegraphics[width=\columnwidth]{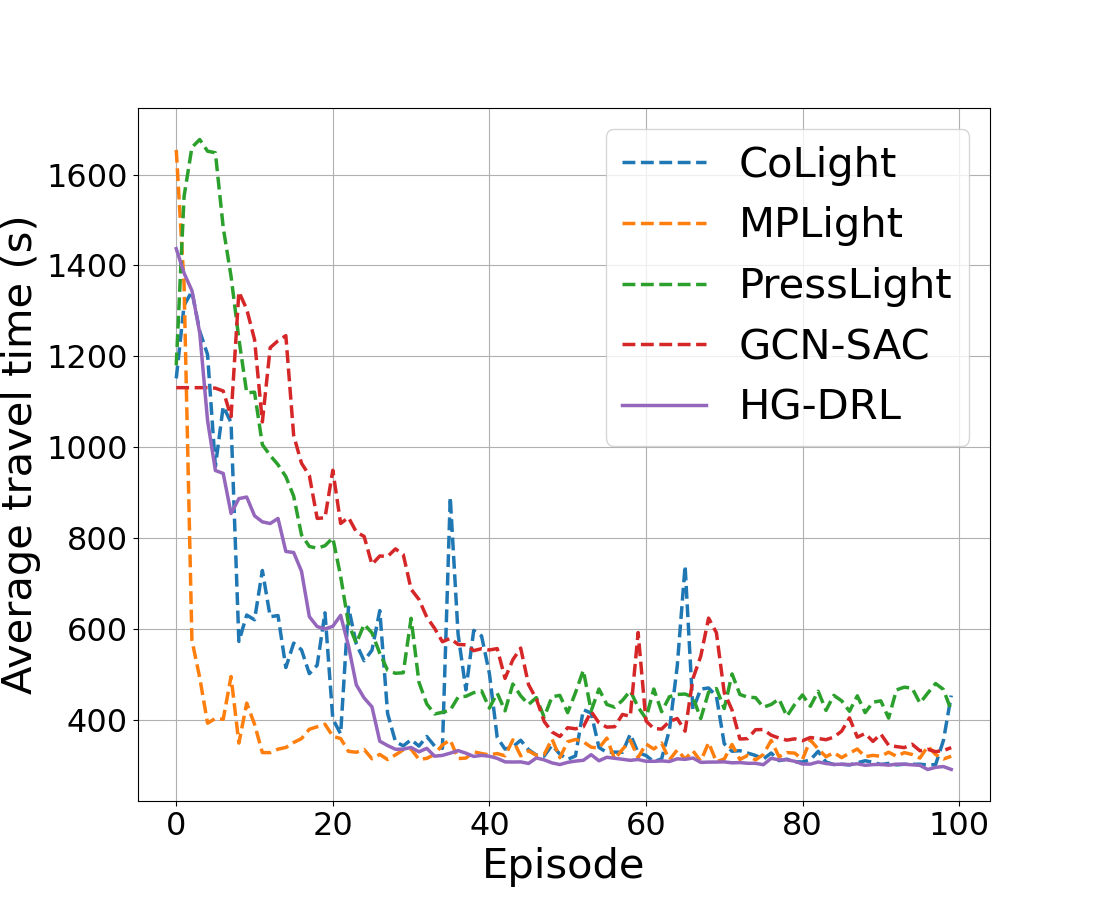}
		\subcaption{$D_\textit{Hangzhou}$}\label{fig:loss_hangzhou}
	\end{minipage}
	\begin{minipage}[b]{0.493\columnwidth}
		\centering
		\includegraphics[width=\columnwidth]{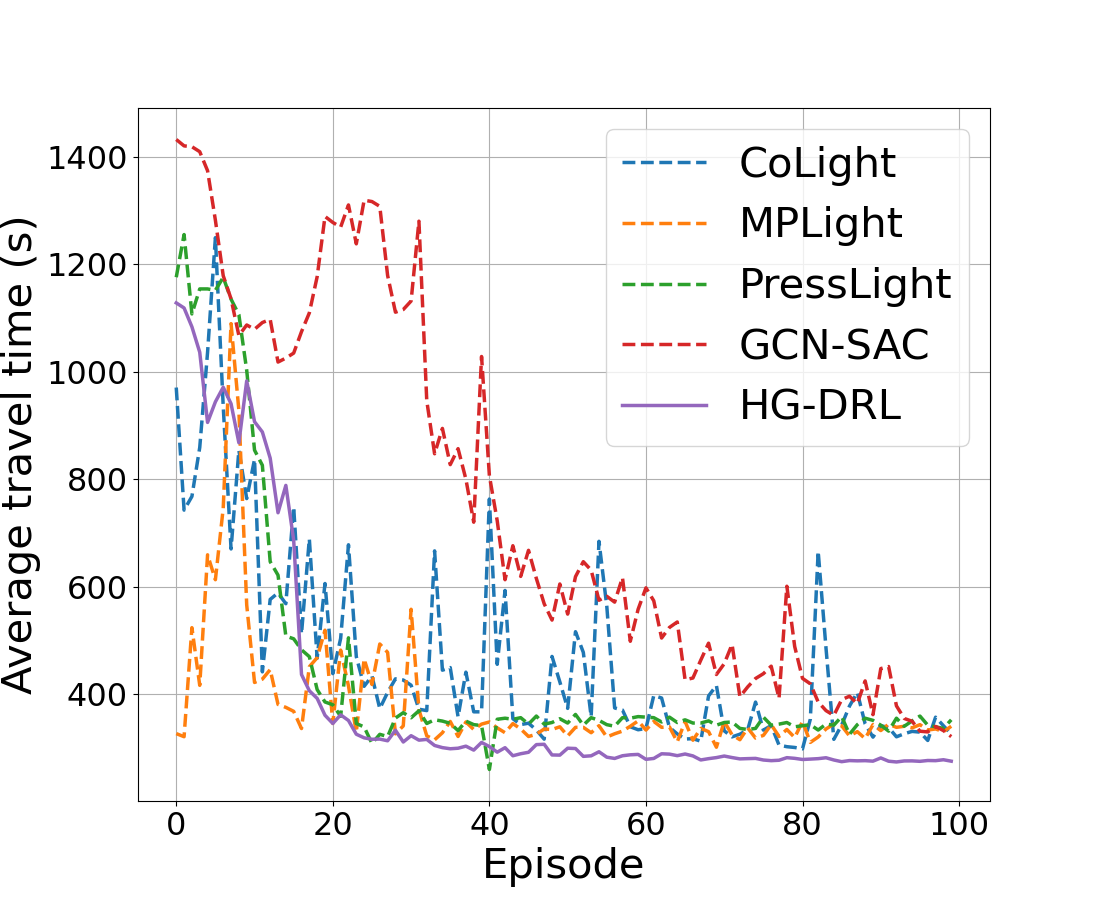}
		\subcaption{$D_\textit{Jinan}$}\label{fig:loss_jinan}
	\end{minipage}
	\caption{\cpy{Learning curves for different methods.}}	
\label{fig:loss}
\end{figure}

Regarding the other comparative methods, the traditional non-reinforcement learning algorithms like Fixed-time and MaxPressure are found to have subpar performance in traffic signal control since they are 
unable to learn from the environmental feedback, and thus cannot adopt more reasonable traffic signal control strategies in real-time based on the overall condition of the road network. Additionally, these two methods aim to enhance the efficiency of ATT at the expense of throughput as shown in Table~\ref{tab:throughput}. On the other hand, reinforcement learning-based methods like PressLight and MPLight utilize a basic DQN in their reinforcement learning module. However, their reward functions solely focus on pressure, which fails to account for the number of vehicles on entering lanes. This oversight results in a decrease in the performance of both average travel time and throughput.

\begin{figure*}[t]
	\centering
	\begin{minipage}[b]{.6666\columnwidth}
		\centering
		\includegraphics[width=\columnwidth]{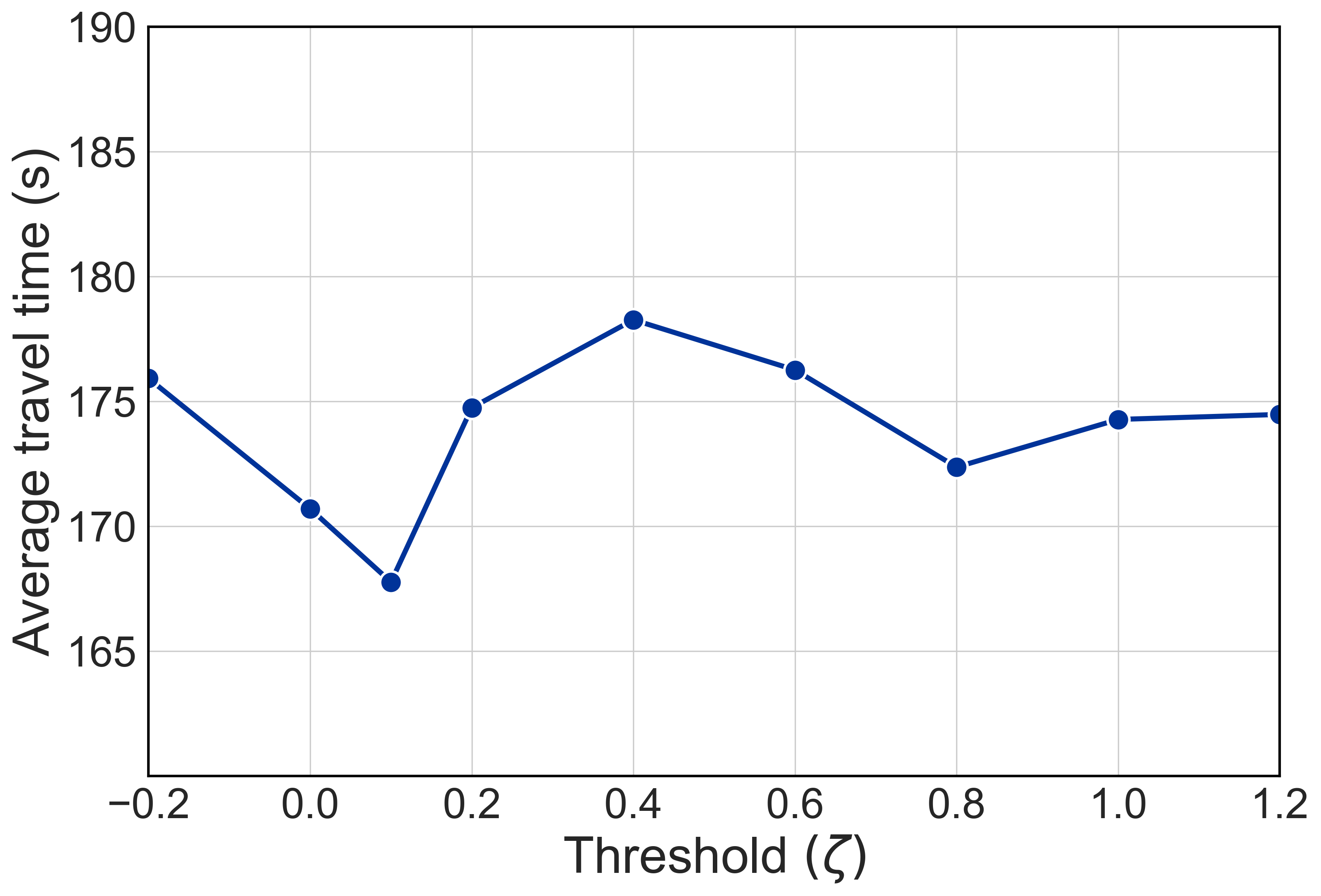}
		\subcaption{$\textit{Bidirect}_\textit{6$\times$6}$}\label{fig:1_zeta}
	\end{minipage}
        \begin{minipage}[b]{.6666\columnwidth}
		\centering
		
		\includegraphics[width=\columnwidth]{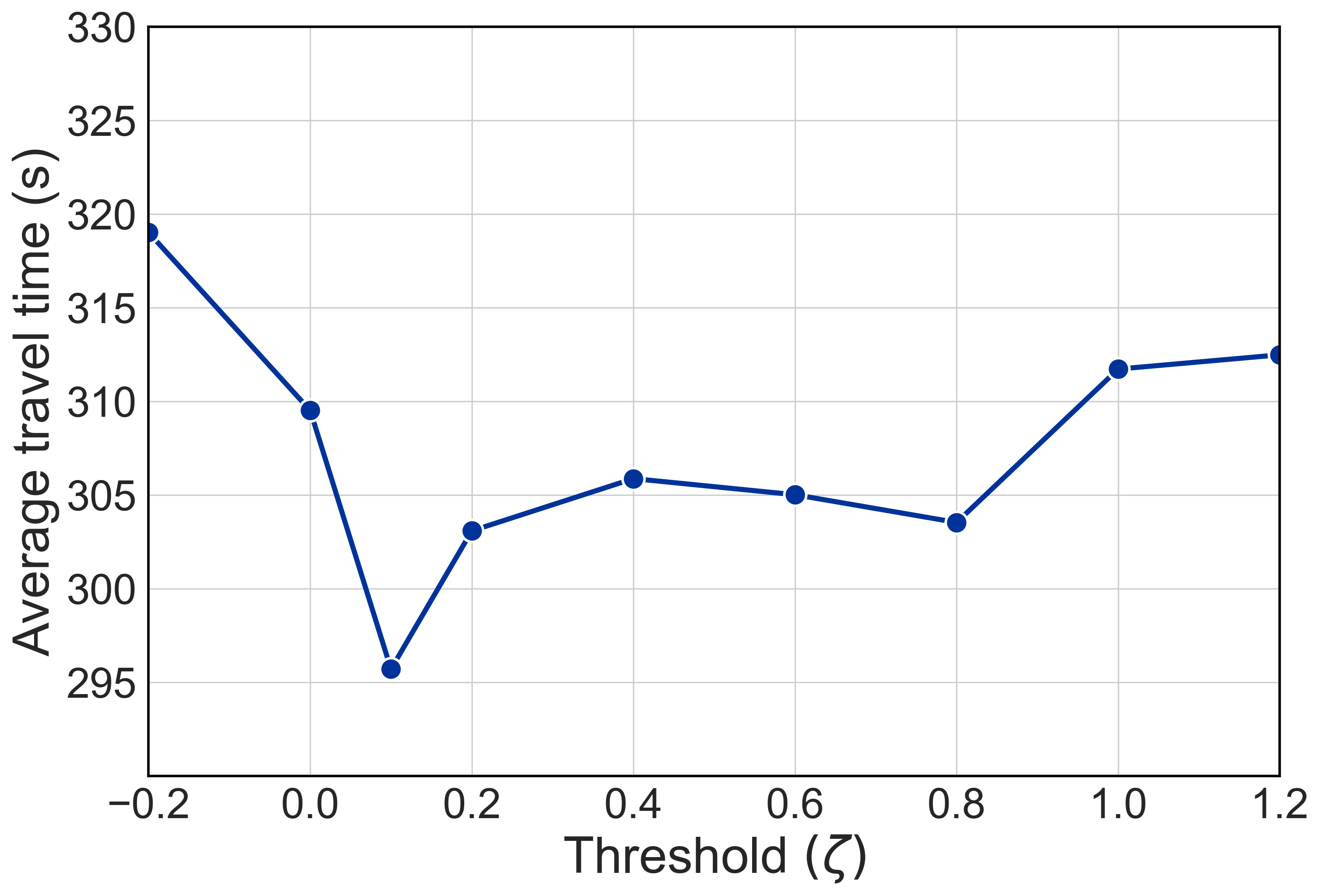}
		\subcaption{$D_\textit{Hangzhou}$}\label{fig:2_zeta}
	\end{minipage}
	\begin{minipage}[b]{.6666\columnwidth}
		\centering
		\includegraphics[width=\columnwidth]{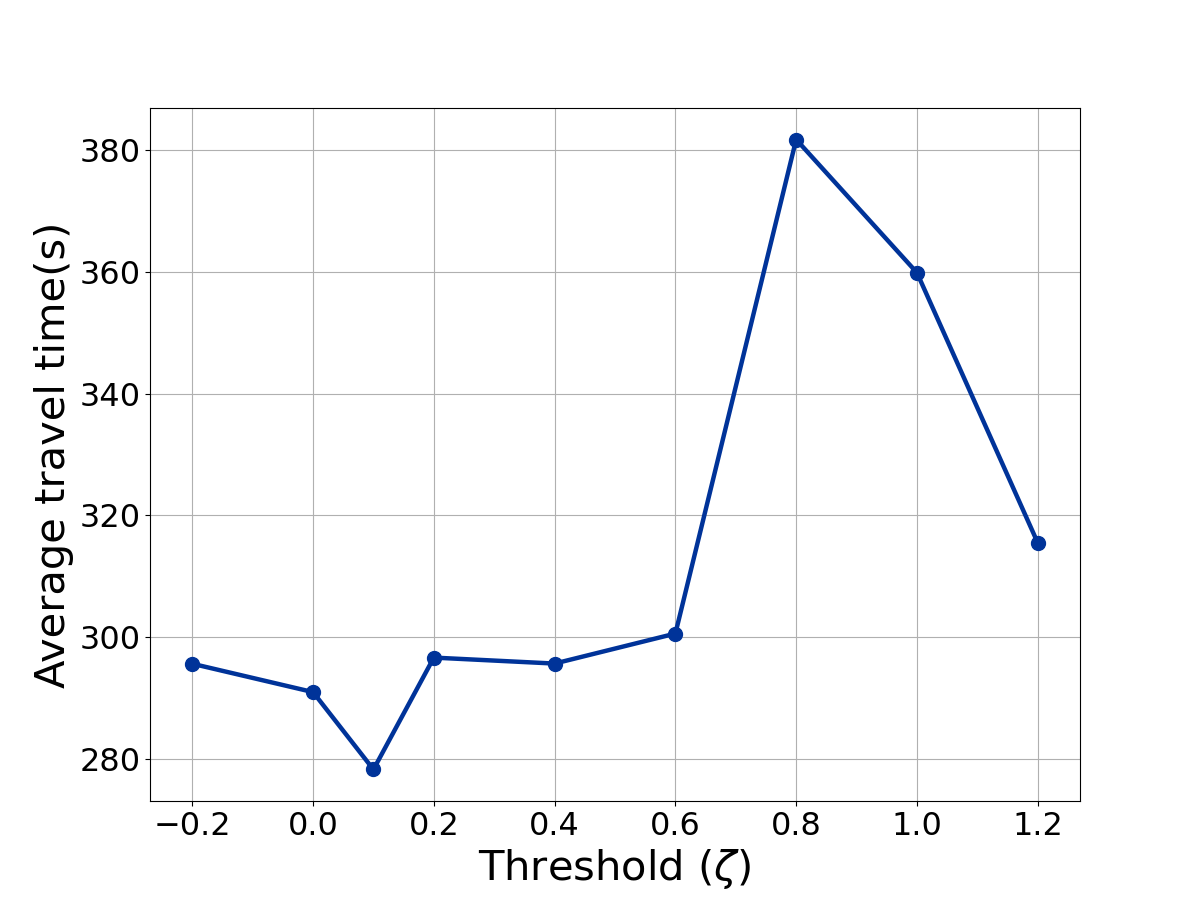}
		\subcaption{$D_\textit{Jinan}$}\label{fig:3_zeta}
	\end{minipage}
		
	\caption{ Impact of threshold $\zeta$.}
	\label{fig:zeta}
\end{figure*}

\begin{figure*}[t]
	\centering
	\begin{minipage}[b]{.666\columnwidth}
		\centering
		\includegraphics[width=\columnwidth]{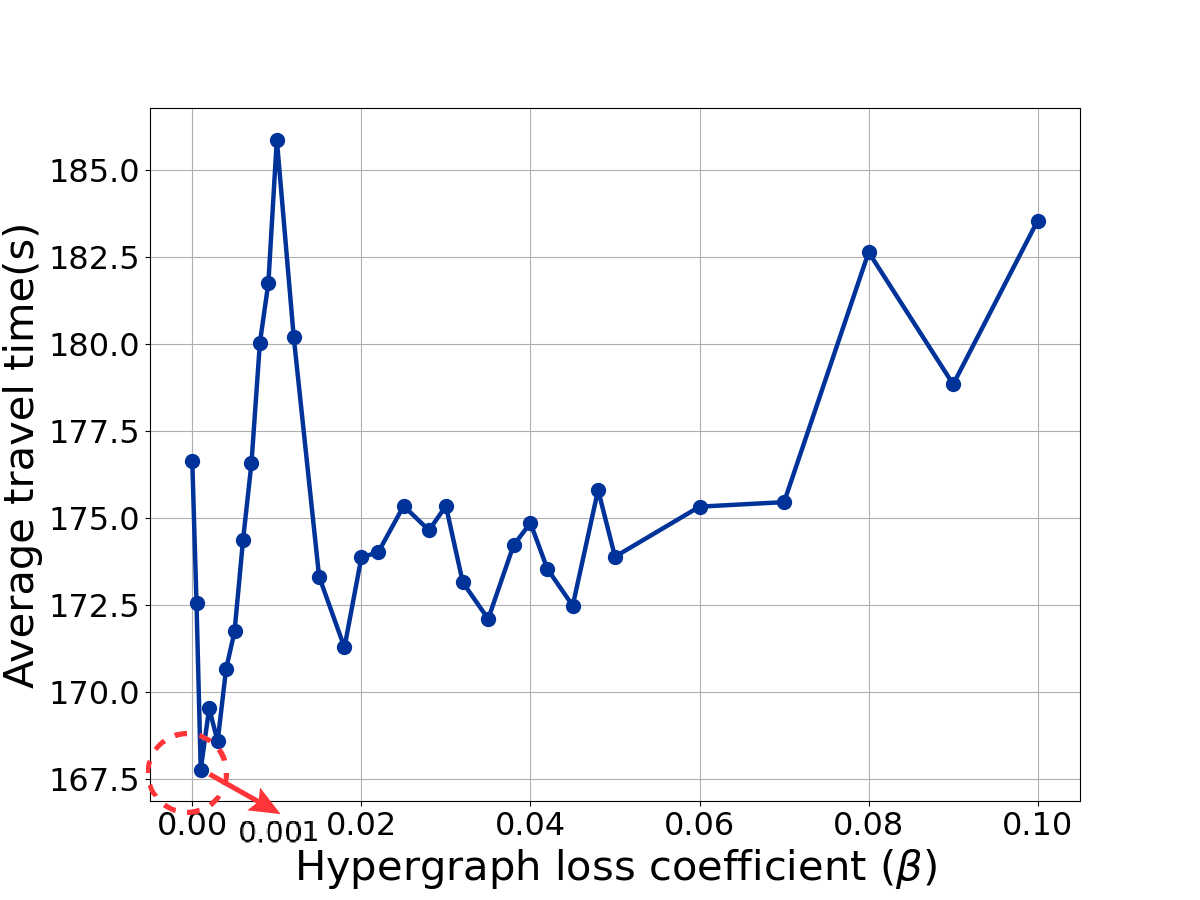}
		\subcaption{$\textit{Bidirect}_\textit{6$\times$6}$}\label{fig:1_beta}
	\end{minipage}
        \begin{minipage}[b]{.666\columnwidth}
		\centering
		
		\includegraphics[width=\columnwidth]{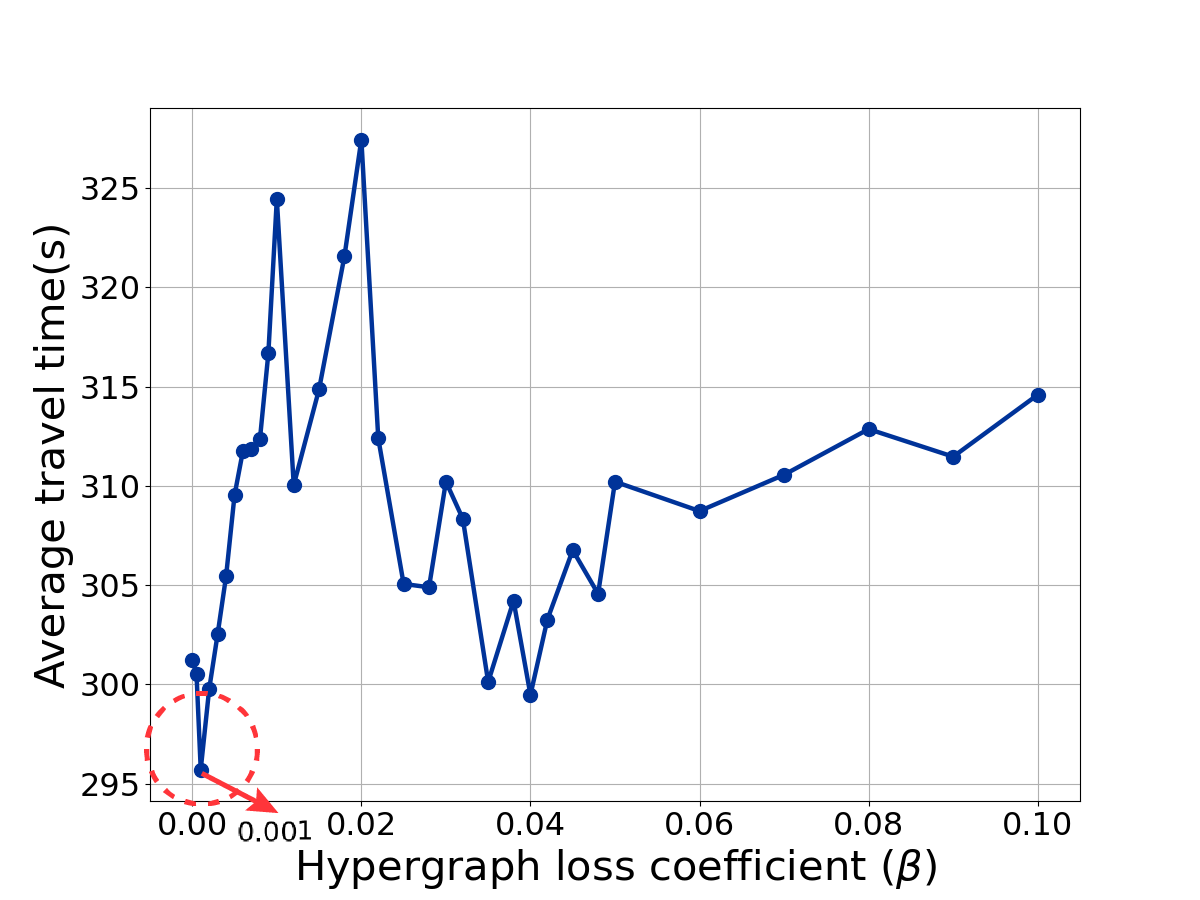}
		\subcaption{$D_\textit{Hangzhou}$}\label{fig:2_beta}
	\end{minipage}
	\begin{minipage}[b]{.666\columnwidth}
		\centering
		\includegraphics[width=\columnwidth]{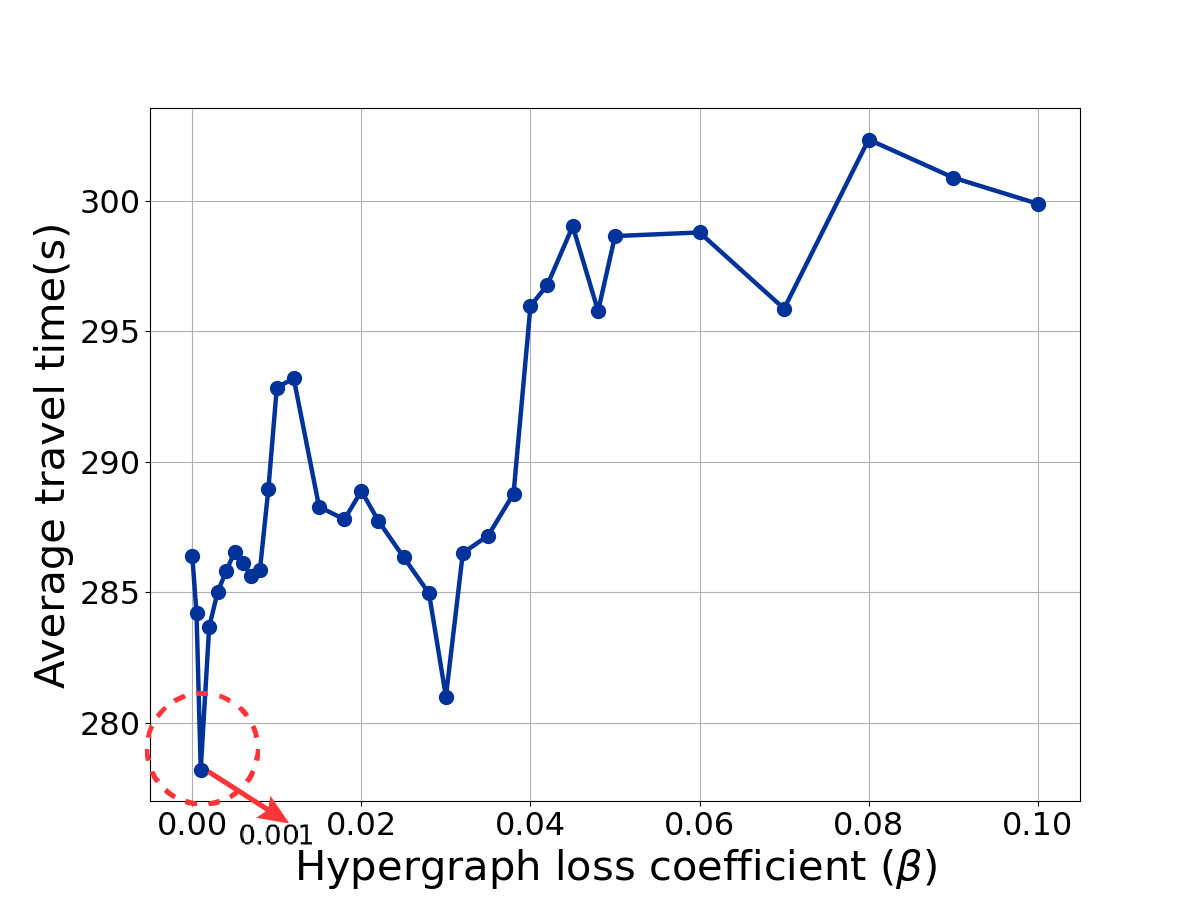}
		\subcaption{$D_\textit{Jinan}$}\label{fig:3_beta}
	\end{minipage}
		
	\caption{ \newnew{Impact of hypergraph loss coefficient $\beta$ }}
	\label{fig:beta}
\end{figure*}

\begin{figure*}[t]
	\centering
	\begin{minipage}[b]{.6666\columnwidth}
		\centering
		\includegraphics[width=\columnwidth]{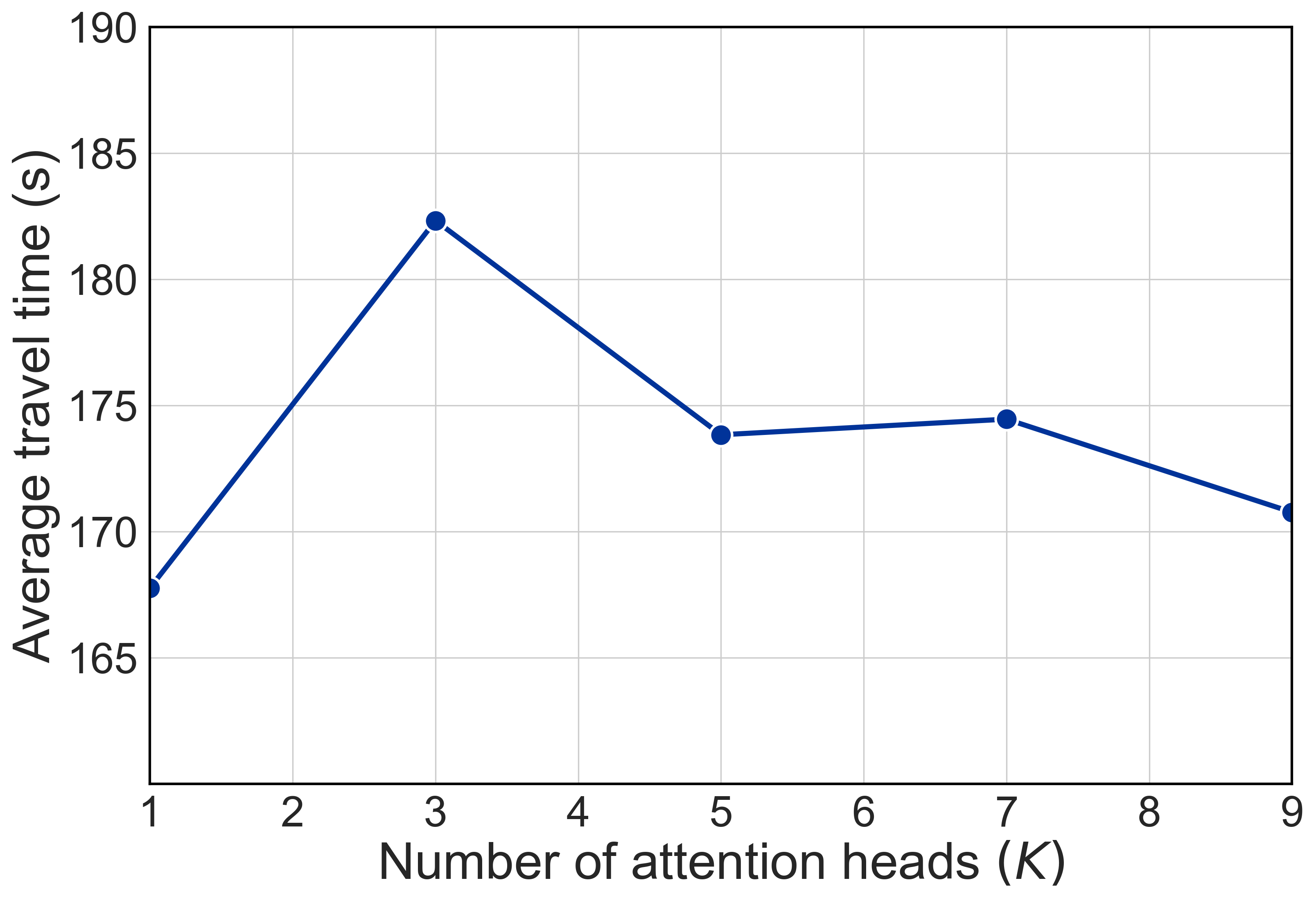}
		\subcaption{$\textit{Bidirect}_\textit{6$\times$6}$}\label{fig:1_att}
	\end{minipage}
        \begin{minipage}[b]{.6666\columnwidth}
		\centering
		
		\includegraphics[width=\columnwidth]{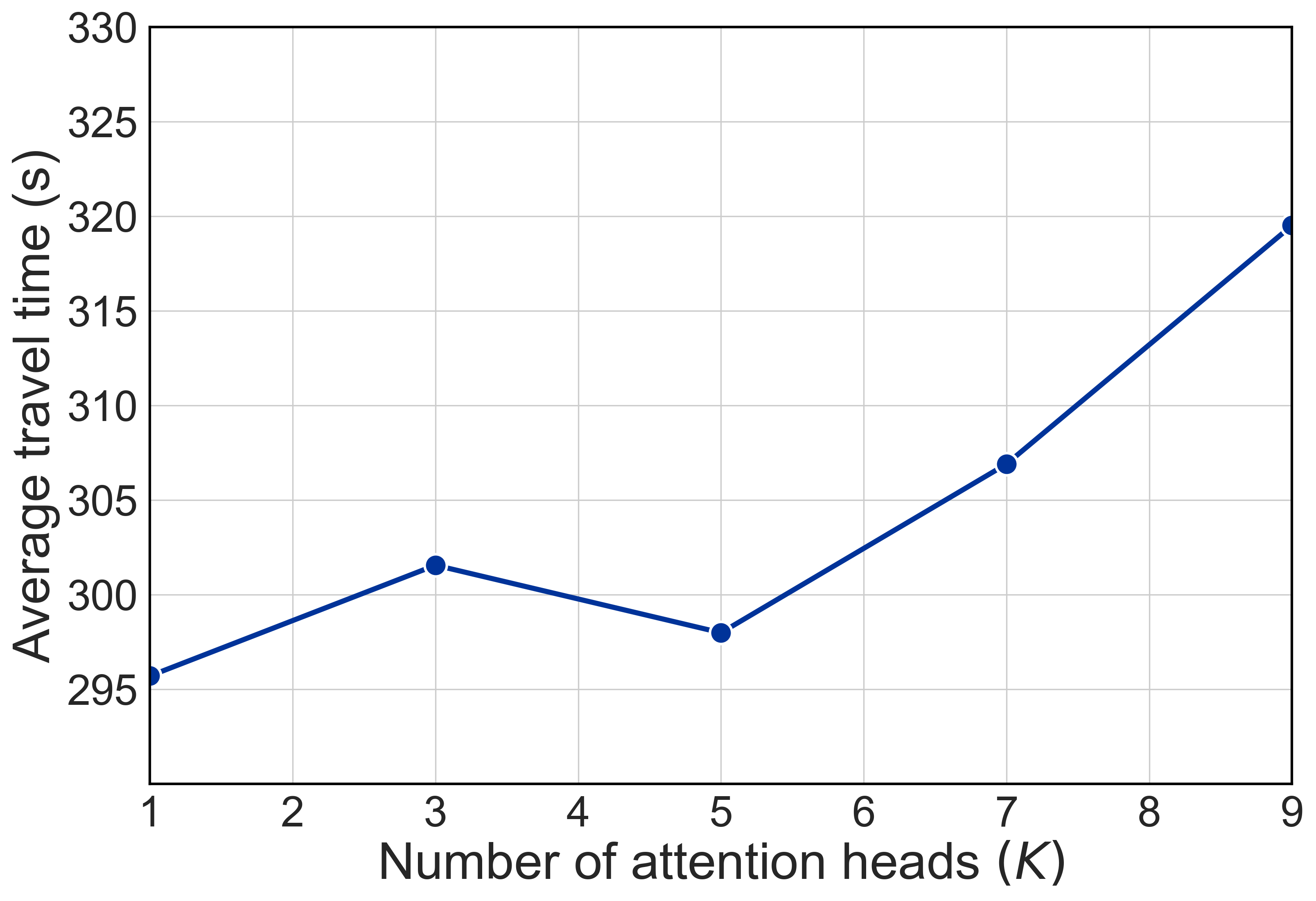}
		\subcaption{$D_\textit{Hangzhou}$}\label{fig:2_att}
	\end{minipage}
	\begin{minipage}[b]{.6666\columnwidth}
		\centering
		\includegraphics[width=\columnwidth]{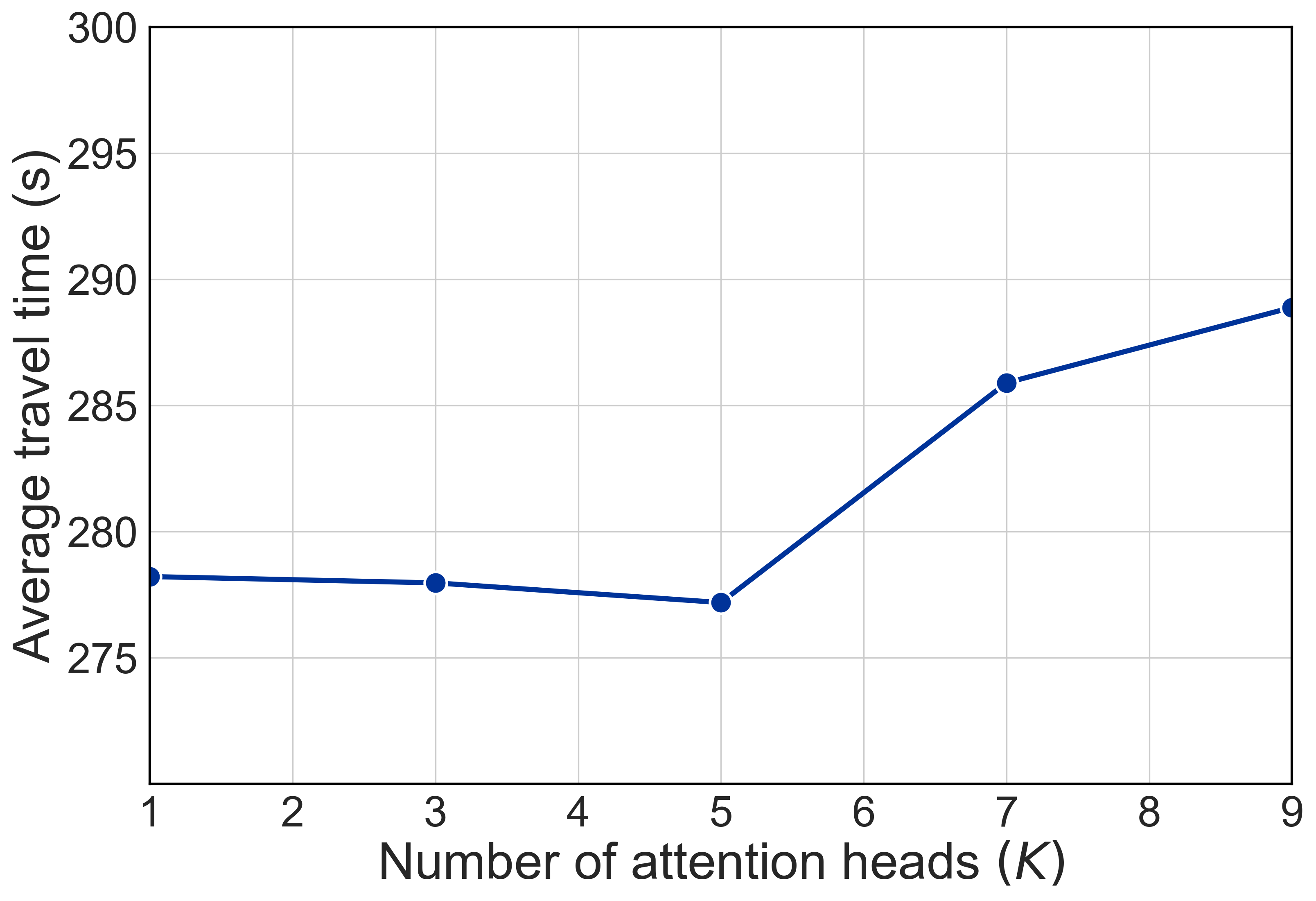}
		\subcaption{$D_\textit{Jinan}$}\label{fig:3_att}
	\end{minipage}
		
	\caption{Impact of attention head number $K$.}
	\label{fig:att}
\end{figure*}

\cpy{
\subsection{Ablation Study}
We conduct the ablation study to validate the importance of hypergraph and DRL across three different datasets as shown in Table~\ref{tab:table1}. In this table, ``w/o HG" indicates the replacement of hypergraph with traditional pairwise GAT, while ``w/o DRL" removes the DRL and installs regular actor-critic method.  The results demonstrate that the hypergraph module enhances the average travel time by at most 16\% compared with the one without it, which verifies that the dynamic construction of spatial and temporal hyperedges effectively captures the spatio-temporal dependencies among traffic signals within road network. Meanwhile, HG-DRL with DRL  can outperform that with actor-critic in terms of travel time up to 10\%. This advantage primarily attributes to the ability of MA-SAC to adaptively learn optimal policies by analyzing the interaction between traffic signal timing and road traffic flow dynamics.}
\begin{table}[ht]
    \centering
    \caption{\cpy{Results of ablation study in terms of average travel time}}
    \label{tab:table1}
    \begin{tabular}{c|ccc}
        \Xhline{1pt}
        Method 
        
         & $D_\textit{Hangzhou}$ & $D_\textit{Jinan}$ & $\textit{Bidirect}_\textit{6$\times$6}$   \\

        \Xcline{1-1}{0.4pt}
        \Xhline{1pt}
        w/o HG  &336.15&332.36 &172.31 \\
        w/o DRL & 305.4 & 310.60&175.4   \\
        \textbf{HG-DRL (Ours)} &\textbf{295.71}&\textbf{278.22} &\textbf{167.77}  \\
 
        \Xhline{1pt}
    \end{tabular} 
    
\end{table}

\subsection{Sensitive Analysis}
We validate the impact of three key parameters of our proposed \algname in three datasets: $\textit{Bidirect}_\textit{6$\times$6}$, $D_\textit{Hangzhou}$, and $D_\textit{Jinan}$.

\subsubsection{Impact of threshold $\zeta$}
Threshold $\zeta$ serves as an important parameter in dynamically constructing hypergraph and capturing crucial neighbor information (See Equation~\ref{eq:zeta}). \figurename~\ref{fig:zeta} show the impact of threshold $\zeta$. With the initial increase in the threshold value $\zeta$, ATT of all vehicles in the entire road network decreases. This reduction occurs due to a reduction in the number of nodes included in the hyperedges constructed by the master node within the road network. For all datasets, the best ATT performance is observed when the threshold $\zeta$ is set at 0.1. If the threshold value $\zeta$ surpasses 0.1, the filtering based on this threshold becomes more stringent, resulting in the exclusion of nodes with potential correlations from information interaction. Consequently, it diminishes the performance of ATT.


\subsubsection{Impact of hypergraph loss coefficient $\beta$}
As an essential component of our proposed method, the hypergraph incorporates two types of hyperedge construction losses into the reinforcement learning loss function (see Equation~\ref{eq:HG_loss}). To investigate the impact of the hypergraph construction loss on the experimental results, we conducted a sensitivity experiment on the hypergraph loss coefficient $\beta$ in the aforementioned three datasets. \newnew{The selected range [0, 0.1] for the $\beta$ value is chosen to effectively scale the two components of the loss function (see Equation 12). This ensures their values fall within a comparable range, facilitating a thorough analysis of the trade-off between them.} The experimental results are shown in \figurename~\ref{fig:beta}. When $\beta$ is set to 0.001, ATT reaches its optimal value across all three datasets. \newinfo{This observation confirms that satisfactory results are not solely dependent on a single $\beta$ value in our proposed method, but rather occur within a consistent range of $\beta$ values.} On the other hand, when $\beta$ is set to 0, it indicates that the loss incurred by hypergraph construction is ignored, resulting in substantial decrease in ATT performance compared to the optimal value. \newnew{While the results demonstrate the promising performance of our proposed method in the current experimental setup, achieving robust sensitivity performance remains a challenging task, which should be further investigated in future work.}

\subsubsection{Impact of Attention Head Number $K$}
To evaluate the effectiveness of multi-head attention, we tested different numbers of attention heads $K$ across three datasets. Selecting an appropriate $K$ is beneficial for better control of intersection signals. As shown in \figurename~\ref{fig:att}, in $\textit{Bidirect}_\textit{6$\times$6}$ and $D_\textit{Hangzhou}$, the ATT increases with the number of attention heads compared to the case when $K=1$. This suggests that having more types of attention do not improve the decision-making performance of model when the head number $K>1$. For $D_\textit{Jinan}$, as the number of attention heads increased, the ATT gradually decreased. However, when $K>5$, the advantage of having more types of attention vanished, leading to a significant increase in ATT.

\vspace{-0.2cm}

\section{Conclusion}\label{sec:conclusion}

In this paper, we propose a framework that can cooperate with multiple edge computing servers for realizing intelligent traffic signal control. Within this framework, we introduce a multi-agent collaborative reinforcement
learning algorithm based on MA-SAC in multi-intersection environments. In order to promote the effective decision-making and improved processing of spatio-temporal information among multiple agents, we incorporate hypergraph learning into the  MA-SAC. We design the dynamic generation of spatio-temporal hyperedges to construct hypergraphs with enhanced modeling capabilities. We then utilize multi-head attention to update intersection information, enabling dynamic spatio-temporal interactions between road intersections. The results obtained from extensive experiments demonstrate that our proposed method outperforms state-of-the-art ones in traffic signal control, achieving improvements in both travel time and throughput, \newinfo{while exhibiting a more stable post-convergence performance}. Our future direction is to study data cleaning and integration schemes to mitigate  the impact of missing or noisy existing in the collected input data, which aims to enhance the robustness of our proposed method.


\bibliographystyle{IEEEtran}
\bibliography{IEEEabrv,ref}

\section*{Bibliographies}
\vskip -2\baselineskip plus -1fil
\begin{IEEEbiography}[{\includegraphics[width=1in,height=1.25in,clip,keepaspectratio]{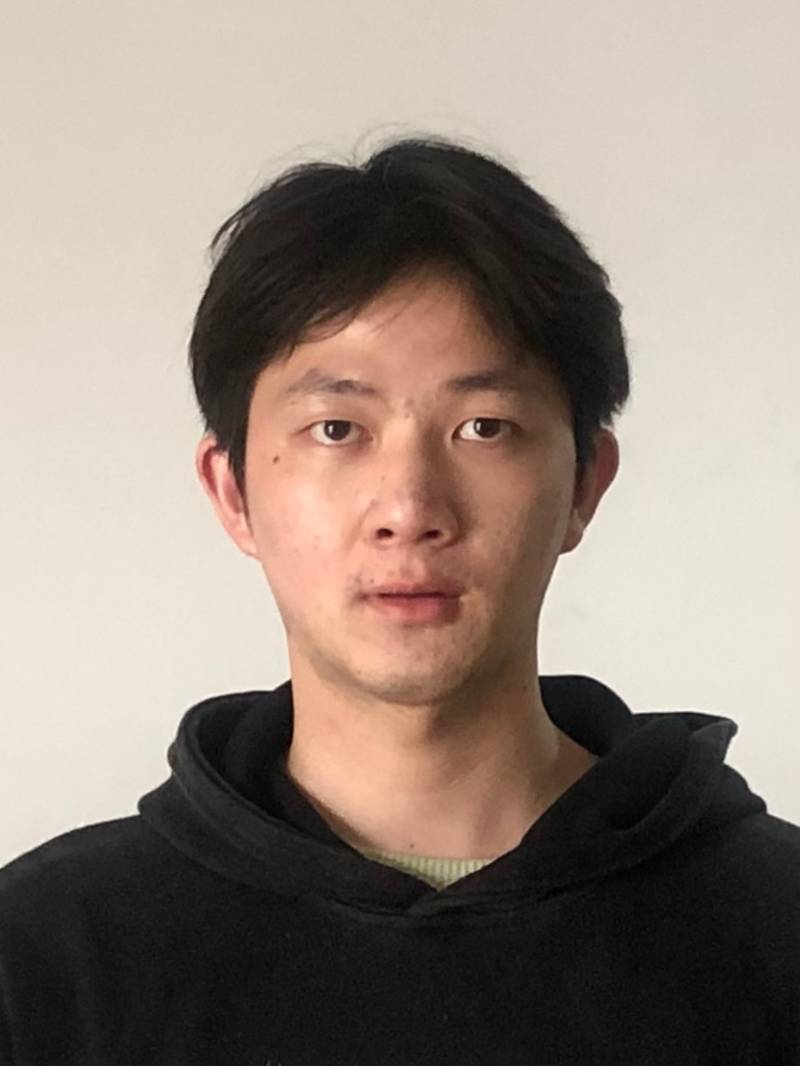}}]{Kang Wang}
received his B.E. degree from the School of  Civil Engineering and Architecture at Wuhan University of Technology, China, in 2022. He is currently working toward a master degree from the School of
Computer Science and Artificial Intelligence at Wuhan
University of Technology, Wuhan, China. His major
interests include edge computing, traffic signal control and graph learning.
\end{IEEEbiography}

\vskip -2\baselineskip plus -1fil

\begin{IEEEbiography}[{\includegraphics[width=1in,height=1.25in,clip,keepaspectratio]{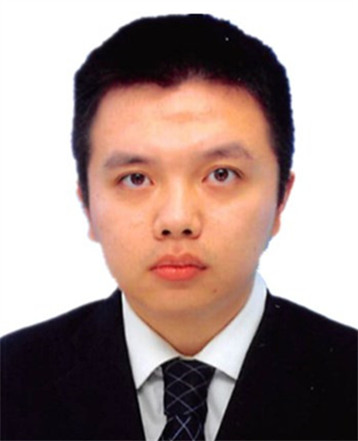}}]{Zhishu Shen}[M'21]
received the B.E. degree from the School of Information Engineering at the Wuhan University of Technology, Wuhan, China, in 2009, and the M.E. and Ph.D. degrees in Electrical and Electronic Engineering and Computer Science from Nagoya University, Japan, in 2012 and 2015, respectively. He is currently an Associate Professor in the School of Computer Science and Artificial Intelligence, Wuhan University of Technology. From 2016 to 2021, he was a research engineer of KDDI Research, Inc., Japan. His major interests include network design and optimization, data learning, edge
computing and the Internet of Things.
\end{IEEEbiography}

\vskip -2\baselineskip plus -1fil

\begin{IEEEbiography}[{\includegraphics[width=1in,height=1.25in,clip,keepaspectratio]{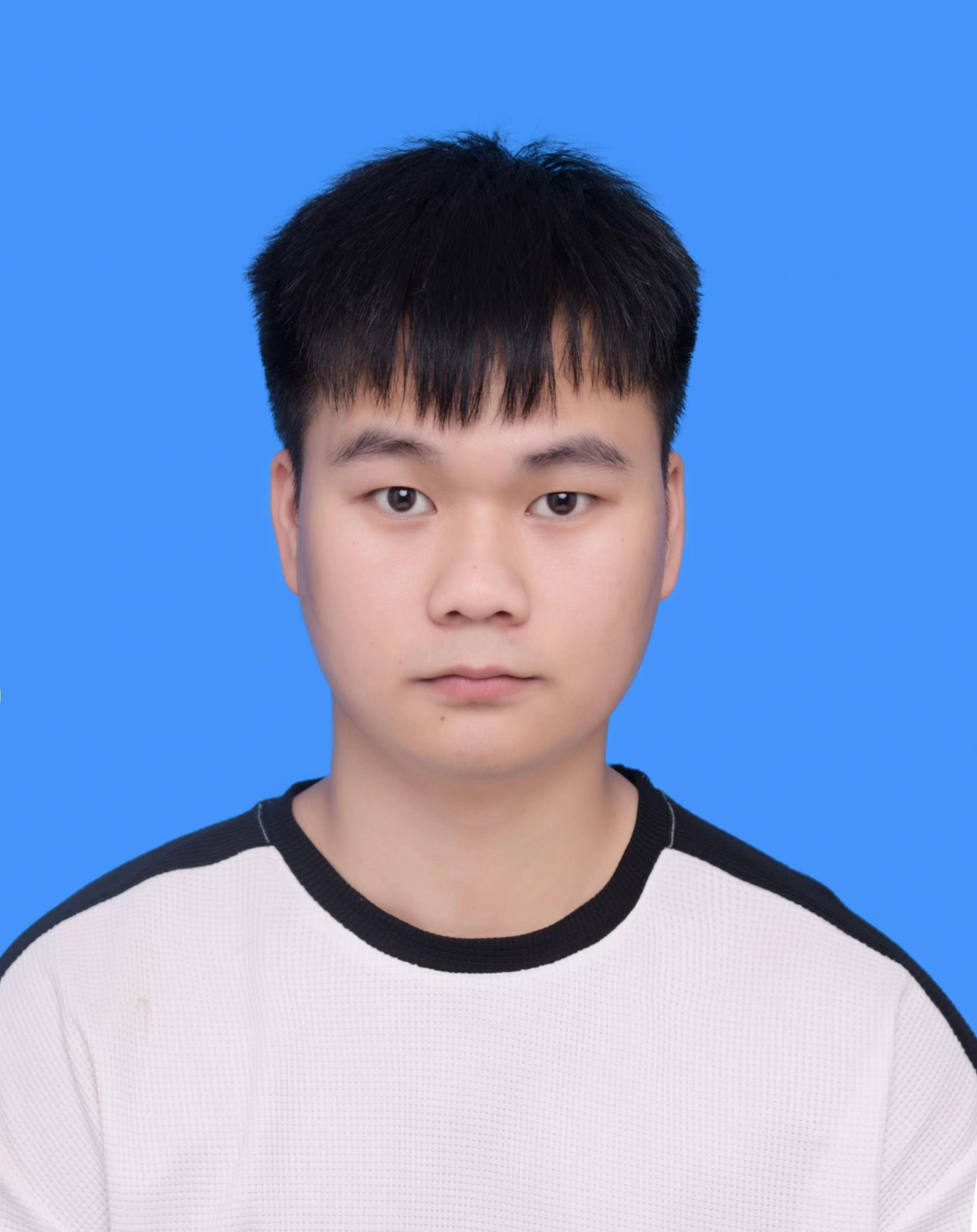}}]{Zhen Lei}
received his B.E. degree from the School of Civil Engineering and Architecture at Wuhan University of Technology, China, in 2022. He is currently working toward a master degree from the School of
Computer Science and Artificial Intelligence at Wuhan
University of Technology, Wuhan, China. His major
interests include intelligent transportation system and machine learning.
\end{IEEEbiography}

\vskip -2\baselineskip plus -1fil

\begin{IEEEbiography}[{\includegraphics[width=1in,height=1.25in,clip,keepaspectratio]{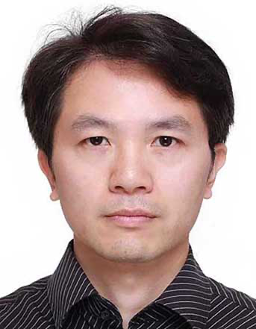}}]{Xianhui Liu}
received his PhD from Tongji University in 2014. He is currently an Associate Professor with the College of Electronic and Information Engineering of Tongji University, Shanghai, China. And he is currently the director of CAD Research Center of Tongji University. He is also a member of Artificial Intelligence Committee of Shanghai. His research interests include industrial internet, in-depth learning and fault prediction.
\end{IEEEbiography}

\vskip -2\baselineskip plus -1fil

\begin{IEEEbiography}[{\includegraphics[width=1in,height=1.25in,clip,keepaspectratio]{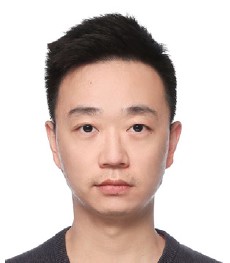}}]{Tiehua Zhang}[M'22]
received B.S. degree from the School of Computer Science and Technology, Jilin University, China, in 2013, M.E. degree from the School of Computing and Information Systems, University of Melbourne, Australia, in 2015, and Ph.D. degree from the School of Software and Electrical Engineering, Swinburne University of Technology, Australia, in 2020. He was a Research Scientist at Ant Group, China from 2021 to 2024,
a Postdoctoral Researcher with the School of Computing, Macquarie University from 2020 to 2021. He is currently an Assistant Professor with the School of Computer Science and Technology, Tongji University, China. His research interests encompass collaborative learning/optimization, edge intelligence, graph learning, and the Internet of Things.
\end{IEEEbiography}

\end{document}